\documentclass[10pt,journal,compsoc]{IEEEtran}

\newif\ifsq
\newif\ifall
\newif\ifconf
\newif\iftr
\newif\iftrcol

\allfalse
\trtrue
\trcolfalse
\conffalse
\sqfalse

\usepackage{etex}

\usepackage[font={bf,sf,scriptsize}]{caption}
\usepackage[font={bf,sf,scriptsize}]{subcaption}

\usepackage{graphicx}
\usepackage{booktabs}
\usepackage{epstopdf}
\usepackage{float}
\usepackage{url}
\usepackage{placeins}
\usepackage{amsmath,amssymb,amsfonts}
\usepackage{mathtools,mathrsfs}
\usepackage{multirow}
\usepackage{rotating}
\usepackage{pbox}
\usepackage[normalem]{ulem}

\usepackage{inconsolata}
\usepackage{listings}

\usepackage[hidelinks]{hyperref}
\usepackage{cleveref}
\usepackage[utf8]{inputenc}
\crefname{section}{§}{§§}
\Crefname{section}{§}{§§}

\usepackage[10pt]{moresize}

\usepackage{xcolor}
\definecolor{darkgrey}{RGB}{70,70,70}
\definecolor{lightgrey}{RGB}{200,200,200}

\usepackage{enumitem}

\usepackage{makecell}

\ifsq
\newcommand{\subparagraph}{}
\usepackage[compact]{titlesec}
\titlespacing*{\section}{0pt}{6pt}{2pt}
\titlespacing*{\subsection}{0pt}{5pt}{1pt}
\titlespacing*{\subsubsection}{0pt}{5pt}{1pt}
\fi

\usepackage[linesnumbered,ruled,vlined]{algorithm2e}
\SetAlFnt{\scriptsize\tt}
\SetAlCapFnt{\scriptsize\sf}
\SetAlCapNameFnt{\scriptsize\sf}

\usepackage{algpseudocode}
\algblock{ParFor}{EndParFor}
\algnewcommand\algorithmicparfor{\textbf{parfor}}
\algnewcommand\algorithmicpardo{\textbf{do}}
\algnewcommand\algorithmicendparfor{\textbf{end\ parfor}}
\algrenewtext{ParFor}[1]{\algorithmicparfor\ #1\ \algorithmicpardo}
\algrenewtext{EndParFor}{\algorithmicendparfor}

\newcommand{\maciej}[1]{\textcolor{blue}{[Maciej: #1]}}
\newcommand{\marcel}[1]{\textcolor{blue}{[Marcel: #1]}}

\newcommand{\macb}[1]{\textbf{\textsf{#1}}}

\usepackage[hang,flushmargin]{footmisc}

\usepackage{bm}

\usepackage{scalerel,stackengine}
\stackMath
\newcommand\rwh[1]{%
\savestack{\tmpbox}{\stretchto{%
  \scaleto{%
      \scalerel*[\widthof{\ensuremath{#1}}]{\kern-.6pt\bigwedge\kern-.6pt}%
          {\rule[-\textheight/2]{1ex}{\textheight}}
            }{\textheight}%
}{0.5ex}}%
\stackon[1pt]{#1}{\tmpbox}%
}

\def\HiLiGA{\leavevmode\rlap{\hbox to \hsize{\color{black!10}\leaders\hrule height 1\baselineskip depth 1ex\hfill}}}
\def\HiLiGB{\leavevmode\rlap{\hbox to \hsize{\color{black!25}\leaders\hrule height 1\baselineskip depth 1ex\hfill}}}
\def\HiLiGC{\leavevmode\rlap{\hbox to \hsize{\color{black!40}\leaders\hrule height 1\baselineskip depth 1ex\hfill}}}
\def\HiLiGD{\leavevmode\rlap{\hbox to \hsize{\color{black!55}\leaders\hrule height 1\baselineskip depth 1ex\hfill}}}
\def\HiLiGE{\leavevmode\rlap{\hbox to \hsize{\color{black!70}\leaders\hrule height 1\baselineskip depth 1ex\hfill}}}
\def\HiLiGF{\leavevmode\rlap{\hbox to \hsize{\color{black!85}\leaders\hrule height 1\baselineskip depth 1ex\hfill}}}

\usepackage{algpseudocode}
\usepackage{tikz}
\usetikzlibrary{calc}
\usepackage{xcolor}
\makeatletter

\usepackage{fontawesome}
\usepackage{pifont}
\usepackage{textcomp}

\usepackage{xcolor}
\usepackage{soul}
\usepackage{dblfloatfix}

\usepackage{units}

\usepackage{colortbl}

\ifconf
\usepackage[firstinits=true,bibencoding=utf8]{biblatex}
\fi

\ifconf

\fi

\newcommand{\noAnswer}{\textcolor{lightgray}{\faQuestionCircle}}

\ifCLASSOPTIONcompsoc
\else
\fi

\ifCLASSINFOpdf
\else
\fi

\usepackage{anyfontsize}

\begin{document}




\title{\vspace{-0.4em}\fontsize{22.5}{27.6}\selectfont High-Performance Routing with Multipathing and Path Diversity in Ethernet and HPC Networks}

\author{Maciej Besta$^1$, Jens Domke$^2$, Marcel Schneider$^1$, Marek Konieczny$^3$,\\ Salvatore Di Girolamo$^1$, Timo Schneider$^1$, Ankit Singla$^1$, Torsten Hoefler$^1$\vspace{0.5em}\\
$^1$Department of Computer Science, ETH Zurich; $^2$RIKEN Center for Computational Science (R-CCS)\\
$^3$Faculty of Computer Science, Electronics and Telecommunications; AGH-UST}

\IEEEtitleabstractindextext{%
\begin{abstract}
The recent line of research into topology design focuses on lowering network
diameter. Many low-diameter topologies such as Slim Fly or Jellyfish that
substantially reduce cost, power consumption, and latency have been proposed. A
key challenge in realizing the benefits of these topologies is \emph{routing}.
On one hand, these networks provide shorter path lengths than established
topologies such as Clos or torus, leading to performance improvements. On the
other hand, the number of shortest paths between each pair of endpoints is much
smaller than in Clos, but there is a large number of non-minimal paths between
router pairs. This hampers or even makes it impossible to use established
multipath routing schemes such as ECMP.
In this work, to facilitate high-performance routing in modern networks, we
analyze existing routing protocols and architectures, focusing on how well they
exploit the diversity of minimal and non-minimal paths. We first develop a
taxonomy of different forms of support for multipathing and overall path
diversity. Then, we analyze how existing routing schemes support this
diversity. Among others, we consider multipathing with both shortest and
non-shortest paths, support for disjoint paths, or enabling adaptivity. To address
the ongoing convergence of HPC and ``Big Data'' domains, we consider routing
protocols developed for both HPC systems and
for data centers as well as general clusters. Thus, we cover architectures and
  protocols based on Ethernet,
InfiniBand, and other HPC networks such as Myrinet. 
Our review will foster developing future high-performance multipathing routing
protocols in supercomputers and data centers.
\end{abstract}

}

\maketitle

\IEEEdisplaynontitleabstractindextext
\IEEEpeerreviewmaketitle

\iftr
{\vspace{-1.0em}\noindent \textbf{This is an extended version of a paper published at\\ IEEE TPDS 2021 under the same title}}
\else
\fi

\iftr
\sethlcolor{white}
\fi

\section{Introduction and Motivation}
\label{sec:intro}

Fat tree~\cite{leiserson1996cm5} and related networks such as
Clos~\cite{clos1953study} are the most commonly deployed topologies in data
centers and supercomputers today, dominating the landscape of Ethernet
clusters~\cite{niranjan2009portland, handley2017re, valadarsky2015}.  However,
many low-diameter topologies such as Slim Fly or Jellyfish that substantially
reduce cost, power consumption, and latency have been proposed. These networks
improve the cost-performance tradeoff compared to fat trees. For instance, Slim
Fly is $\approx$$2\times$ more cost- and power-efficient at scale than fat
trees, simultaneously delivering $\approx$25\% lower
latency~\cite{besta2014slim}. 
%

\begin{figure}[h]
\ifsq\vspace{-1em}\fi
\centering
\includegraphics[width=1.0\columnwidth]{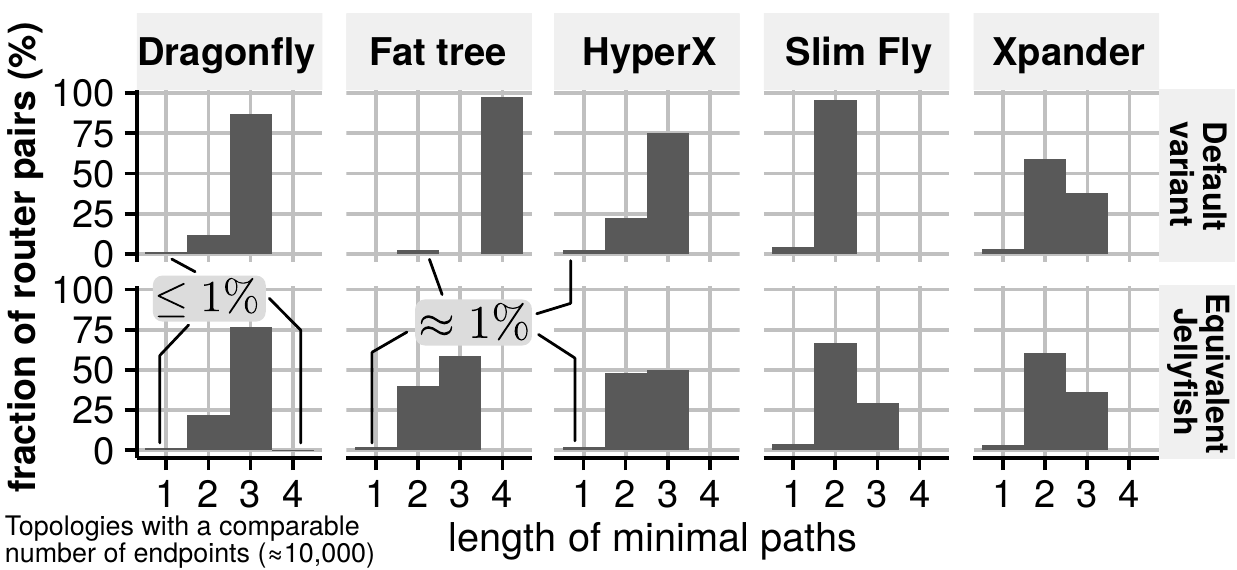}\\
\includegraphics[width=1.0\columnwidth]{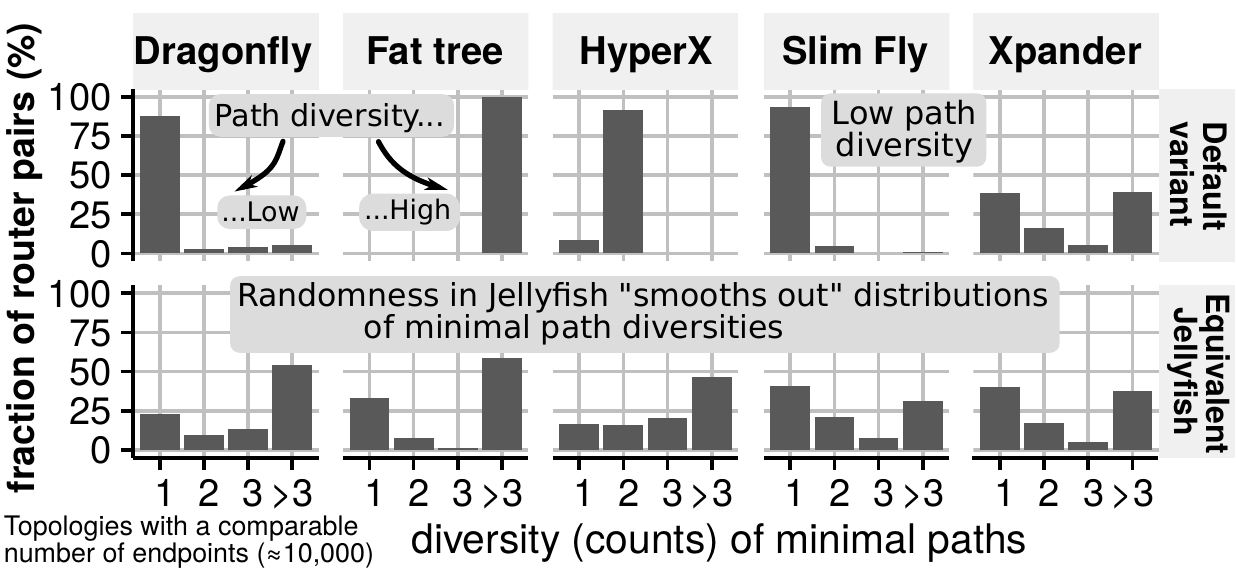}
\caption{\textmd{Distributions of lengths and counts of shortest paths in
low-diameter topologies and in fat trees. When analyzing counts of minimal
paths between a router pair, we consider disjoint paths (no shared
links). An equivalent Jellyfish network is constructed using the
same number of identical routers as in the corresponding non-random topology (a
plot taken from our past work~\cite{besta2019fatpaths}).}}
\ifsq\vspace{-2em}\fi
\label{fig:shortest_path_length}
\label{fig:shortest_path_multiplicity}
\end{figure}

A key challenge in realizing the benefits of these topologies is routing. On
one hand, due to their lower diameters, these networks provide shorter path
lengths than fat trees and other traditional topologies such as torus. However,
as illustrated by our recent research efforts~\cite{besta2019fatpaths},
\emph{the number of shortest paths between each pair of endpoints is much
smaller than in fat trees}. Selected results are illustrated in
Figure~\ref{fig:shortest_path_multiplicity}. In this figure, we compare
established three-level fat trees (FT3) with representative modern low-diameter
networks: {Slim Fly} (SF)~\cite{besta2014slim, besta2018slim} (a variant with
diameter~2), {Dragonfly} (DF)~\cite{kim2008technology} (the ``balanced''
variant with diameter~3), {Jellyfish} (JF)~\cite{singla2012jellyfish} (with
diameter~3), {Xpander} (XP)~\cite{valadarsky2015} (with diameter~$\le 3$), and
{HyperX} (Hamming graph) (HX)~\cite{ahn2009hyperx} that generalizes Flattened
Butterflies (FBF)~\cite{kim2007flattened} with diameter~3. As
observed~\cite{besta2019fatpaths}, \emph{``in DF and SF, most routers are
connected with \emph{one} minimal path. In XP, more than 30\% of routers are
connected with \emph{one} minimal path.''} In the corresponding JF networks
(i.e., random Jellyfish networks constructed using the same number of identical
routers as in the corresponding non-random topology), \emph{``the results are
more leveled out, but pairs of routers with one shortest part in-between still
form large fractions.  FT3 and HX show the highest diversity.''} We conclude
that in all the considered low-diameter topologies, \emph{\textbf{shortest
paths fall short:}} at least a large fraction of router pairs are connected
with \emph{\textbf{only one}} shortest path.

\ifconf
\fi

Simultaneously, \textbf{these low-diameter topologies offer high diversity of
\emph{non-minimal paths}}~\cite{besta2019fatpaths}{. They} provide at least three
disjoint ``almost''-minimal paths (i.e., paths that are one hop longer than
their corresponding shortest paths) per router pair (for the majority of
pairs). For example, in Slim Fly (that has the diameter of~2), 99\% of router
pairs are connected with multiple non-minimal paths of
length~3~\cite{besta2019fatpaths}.

The above properties of low-diameter networks place unprecedented design
challenges for performance-conscious routing protocols. First, as shortest
paths fall short, one must resort to non-minimal routing, which is usually more
complex than the minimal one. Moreover, as topologies lower their diameter,
their link count is also reduced. Thus, even if they do indeed offer more than
one non-minimal path between pairs of routers, the corresponding routing
protocol must carefully use these paths in order not to congest the network
(i.e., the path diversity is still a scarce resource demanding careful
examination and use). Third, a shortage of shortest paths means that one cannot
use established multipath routing\footnote{{\textbf{Multipath}
routing indicates a routing protocol that uses more than one
path in the network, for at least one pair of communicating endpoints. We
consider multipathing both within a single flow/message (e.g., as in
spraying single packets across multiple paths, cf.~\mbox{\cref{sec:ps}}), and
multipath across flows/messages (e.g., as in standard ECMP, where different
flows follow different paths~\mbox{\cref{sec:ecmp}}). \textbf{Path diversity}
indicates whether a given network topology offers multiple
paths between different routers (i.e., has potential for speedups from multipath routing).}} schemes such as
Equal-Cost Multi-Path (ECMP)~\cite{hopps2000analysis}, which usually assume
that \emph{different paths between communicating entities are minimal and have
equal lengths}.  Restricting traffic to these paths does not utilize the path
diversity of low-diameter networks.

In this work, to facilitate overcoming these challenges and to propel designing 
high-performance routing for modern interconnects, we develop a
taxonomy of different forms of support for path diversity by a
routing design. These forms of support include (1) enabling multipathing
using both (2) shortest and (3) non-shortest paths, (4) explicit consideration
of disjoint paths, (5) support for adaptive load balancing across these paths,
and (6) genericness (i.e., being applicable to different topologies).
We also discuss additional aspects, for example whether a given design 
uses multipathing to enhance its \emph{resilience}, \emph{performance}, or both.

Then, we use this taxonomy to categorize and analyze a wide selection of
existing routing designs. Here, we consider two fundamental classes of routing
designs: simple routing \emph{building blocks} (e.g., 
ECMP~\cite{hopps2000analysis} or Network Address Aliasing (NAA)) and routing
\emph{architectures} (e.g., PortLand~\cite{niranjan2009portland} or
PARX~\cite{domke_hyperx_2019}).
While analyzing respective routing architectures, we include and investigate the
architectural and technological details of these designs, for example whether a
given scheme is based on the simple Ethernet architecture, the full TCP/IP
stack, the InfiniBand (IB) stack, or other HPC designs. This enables network architects and
protocol designers to gain insights into supporting path diversity in the
presence of different technological constraints.
 
\ifconf
\fi

\begin{figure}[h!]
\ifsq\vspace{-1em}\fi
\centering
\ifconf
\includegraphics[width=0.24\textwidth]{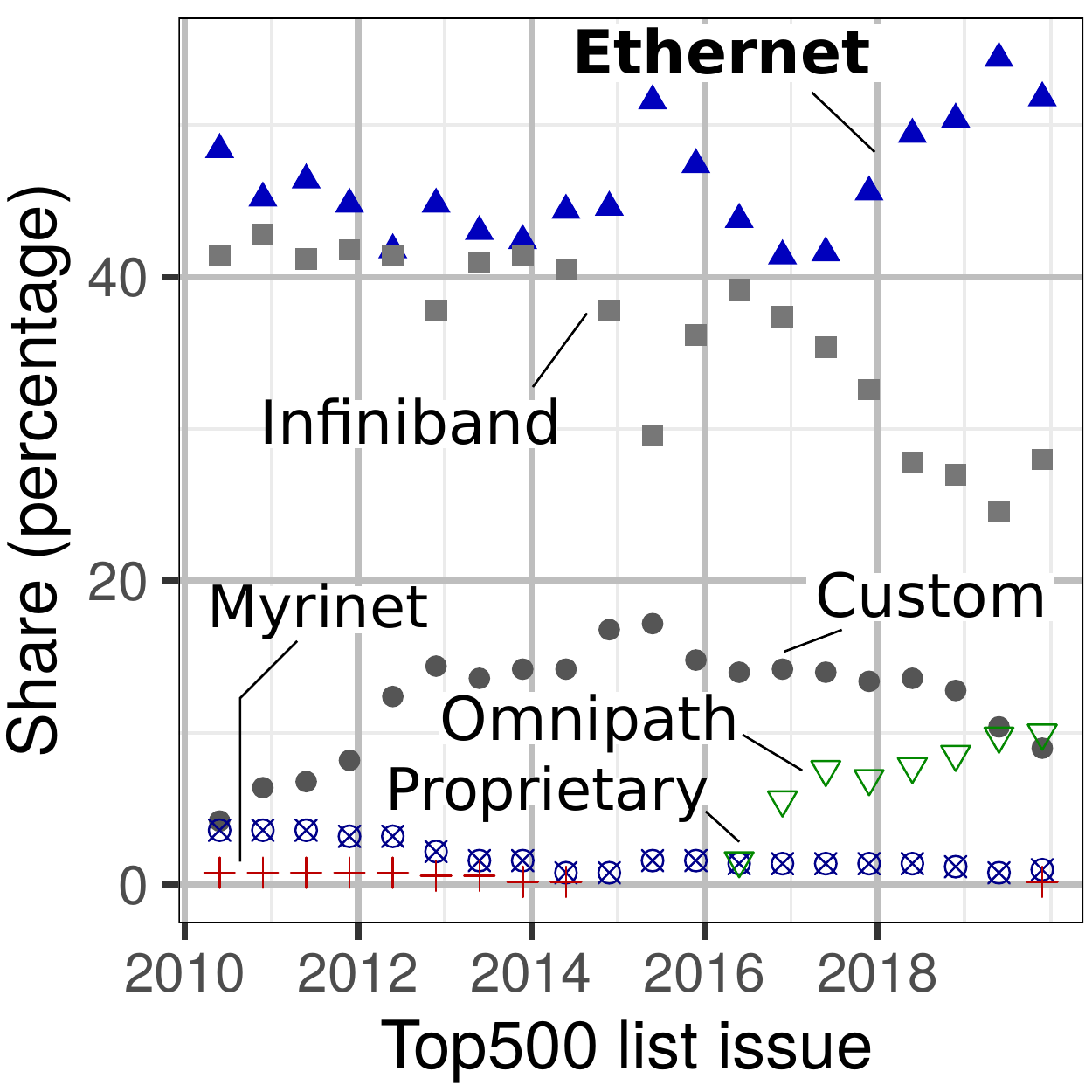}
\fi
\iftr
\includegraphics[width=0.34\textwidth]{top500_2020.pdf}
\fi
\ifsq\vspace{-0.5em}\fi
\caption{\textmd{The share of different interconnect technologies in the Top500
systems (a plot taken from our past work~\cite{besta2019fatpaths}).}}
\ifsq\vspace{-0.75em}\fi
\label{fig:motivation}
\end{figure}

We consider protocols and architectures that originated in both the
HPC and data center as well as general networking 
communities. This is because all these environments are important in
today's large-scale networking landscape. While the most powerful Top500
systems use vendor-specific or InfiniBand (IB) interconnects, more than half of
the Top500 (e.g., in the June 2019 or in the November 2019 issues)
machines~\cite{dongarra1997top500} are based on Ethernet, see
Figure~\ref{fig:motivation}. We observe similar numbers for the Green500 list.
The importance of Ethernet is increased by the \emph{``convergence of HPC and
Big Data''}, with cloud providers and data center operators aggressively aiming
for high-bandwidth and low-latency fabrics~\cite{valadarsky2015, handley2017re,
vanini2017letflow}. 
\iftr
Another example is Mellanox, with its Ethernet sales for
the 3rd quarter of 2017 being higher than those for
IB~\cite{mellanox-sales}. 
Similar trends are observed in more recent
numbers~\mbox{\cite{mellanox-sales-2020}}.
\fi
\ifconf
{Another example is Mellanox, with its Ethernet sales
being higher than those for
InfiniBand in recent years~\mbox{\cite{mellanox-sales}}.}
\fi
\iftr
Similar trends are observed in more recent numbers: \emph{``Sales of Ethernet
adapter products increased 112\% year-over-year (...) we are shipping 400 Gbps
Ethernet switches''}~\cite{mellanox-sales-2020}.
\fi
At the same time, {IB's sales have been growing by 27\%
year-over-year~\mbox{\cite{mellanox-sales-2020}}}.
\iftr
This is ``led by strong demand for the HDR 200 gigabit
solutions''~\cite{mellanox-sales-2020}.
\fi
Thus, our analysis can facilitate developing multipath routing in both
IB-based supercomputers but also in a broad landscape of cloud computing
infrastructure such as data
centers.

\ifconf
\fi

\begin{figure*}[t]
\ifsq\vspace{-1.0em}\fi
\centering
\includegraphics[width=1.0\textwidth]{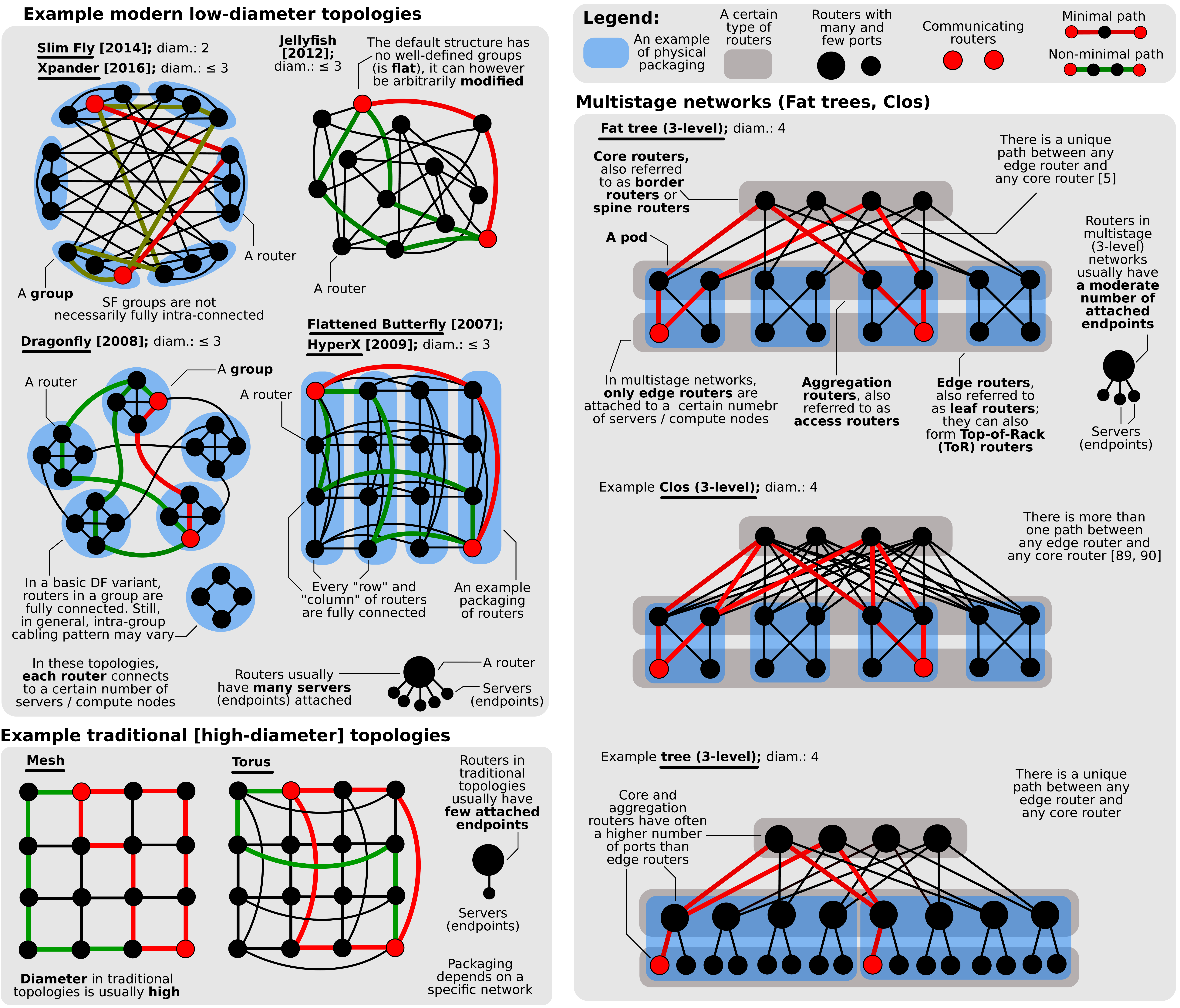}
\vspace{-1.5em}
\caption{\textbf{Illustration of network topologies}
related to the routing protocols and schemes considered in this work.
\textcolor{red}{Red color} indicates an example shortest path between routers.
\textcolor{green}{Green color} indicates example alternative non-minimal paths.
\textcolor{blue}{Blue color} illustrates grouping of routers.
}
\ifsq\vspace{-1em}\fi
\label{fig:topologies}
\end{figure*}

\iftr
\vspace{0.5em}
In general, we provide the following contributions:

\begin{itemize}[noitemsep, leftmargin=0.5em]
\item We provide the first taxonomy of networking architectures and the
associated routing protocols, focusing on the offered
support for path diversity and multipathing.
\item We use our taxonomy to categorize a wide selection of routing
designs for data centers and supercomputers.
\item We investigate the relationships between support for path diversity and
architectural and technological details of different routing protocols. 
\item We discuss in detail the design of representative protocols.
\item We are the first to analyze multipathing schemes related to both
supercomputers and the High-Performance Computing (HPC) community (e.g., the
Infiniband stack) and to data centers (e.g., the TCP/IP stack).
\end{itemize}
\fi

\vspace{0.5em}
\macb{Complementary Analyses}
There exist surveys on multipathing~\cite{radi2012multipath, al2017survey,
zin2015survey, anasane2016survey, tsai2006review, li2016multipath,
singh2015survey}. Yet, none focuses on multipathing and path diversity offered
by routing in data centers or supercomputers. For example, Lee and
Choi describe multipathing in the general Internet and telecommunication
networks~\cite{lee2002survey}.  Li et al.~\cite{li2016multipath} also focus on
the general Internet, covering aspects of multipath transmission related to all
TCP/IP stack layers. Singh et al.~\cite{singh2015survey} cover only
a few multipath routing schemes used in data centers, focusing on a broad
Internet setting.  Moreover, some works are dedicated to performance
evaluations of a few schemes for multipathing~\cite{aggarwal2016performance,
huang2009performance}. Next, different works are dedicated to multipathing
in sensor networks~\cite{anasane2016survey, al2017survey, radi2012multipath,
zin2015survey}.  
Finally, there are analyses of other aspects of data center networking, for
example energy efficiency~\cite{shuja2014survey, beloglazov2011taxonomy},
optical interconnects~\cite{kachris2012survey}, network
virtualization~\cite{jain2013network, bari2012data}, overall
routing~\cite{chen2011survey}, general data center networking with focus on
traffic control in TCP~\cite{noormohammadpour2017datacenter}, low-latency data
centers~\cite{liu2013low}, the TCP incast problem~\cite{ren2014survey},
bandwidth allocation~\cite{chen2014allocating}, transport
control~\cite{zhang2013survey}, general data center networking for
clouds~\cite{wang2015survey}, congestion
management~\cite{joglekar2016managing}, reconfigurable
data center networks~\cite{foerster2019survey}, and transport
protocols~\cite{sreekumari2016transport, polese2019survey}.
We complement all these works, focusing solely on \emph{multipath \ul{routing}
in supercomputers, data centers, and small clusters.}
%
%
As opposed to other works with broad
focus, \emph{we specifically target the \ul{performance} aspects of
multipathing and path diversity}. Our survey is the first to deliver a
taxonomy of the path diversity features of routing schemes, to categorize
existing routing protocols based on this taxonomy, and to consider both
traditional TCP/IP and Ethernet designs, but also protocols and concepts
traditionally associated with HPC, for example multipathing in
InfiniBand~\cite{pfister2001introduction, fompi-paper}.

\section{Fundamental Notions}

We first outline fundamental notions: \emph{network topologies},
\emph{network stacks}, and associated \emph{routing concepts} and \emph{designs}.
%


While we do not conduct any theoretical investigation, we state -- for clarity
-- a network model used implicitly in this work.  We model an
interconnection network as an undirected graph $G = (V,E)$; $V$ and $E$ are
sets of routers, also referred to as nodes ($|V| = N_r$), and full-duplex
inter-router physical links.  Endpoints (also referred to as servers or compute
nodes) are \emph{not} modeled explicitly.

\subsection{Network Topologies}\label{sec:nettopo}

We consider routing in different network topologies. The most important
associated topologies are in Figure~\ref{fig:topologies}.  We only briefly
describe their structure that is used by routing architectures to enable
multipathing (a detailed analysis of respective topologies in terms of their
path diversity is available elsewhere~\cite{besta2019fatpaths}).
In most networks, routers form \emph{groups} that are intra-connected with the
same pattern of cables.  We indicate such groups with the blue
color.

Many routing designs are related to \textbf{fat trees (FT)}~\cite{leiserson1996cm5}
and \textbf{Clos (CL)}~\cite{clos1953study}. In these networks (broadly referred to
as ``\textbf{multistage topologies (MS)}''), a certain fraction of routers is attached to
endpoints while the remaining routers are only dedicated to forwarding traffic.
A common realization of these networks consists of three stages (layers) of
routers: \emph{edge} (\emph{leaf}) routers, \emph{aggregation} (\emph{access})
routers, and \emph{core} (\emph{spine, border}) routers. 
\ifconf
\fi
%
{Edge and aggregation routers are additionally grouped into \emph{pods},
to facilitate physical layout (cf.~Fig.~\mbox{\ref{fig:topologies}}).}
Only edge routers
connect to endpoints. Aggregation and core routers only forward traffic; they
enable multipathing. 
The exact form of multipathing depends on the topology variant. Consider a pair
of communicating edge routers (located in different pods/groups).
In fat trees, multipathing is enabled by selecting \emph{different core
routers} and \emph{different aggregation routers} to forward traffic between
the same communicating pair of edge routers. Importantly, after fixing the
core router, \emph{there is a unique path between the communicating edge routers}.
In Clos, in addition to such multipathing enabled by selecting different core
routers, \emph{one can also use different paths between a specific edge and
core router}. Finally, simple trees are similar to fat trees in that
fixing different core routers enables multipathing; still, \emph{one cannot
multipath by using different aggregation routers}.


The most important \textbf{modern low-diameter networks} are {Slim Fly}
(SF)~\cite{besta2014slim}, {Dragonfly} (DF)~\cite{kim2008technology},
{Jellyfish} (JF)~\cite{singla2012jellyfish}, {Xpander}
(XP)~\cite{valadarsky2015}, and {HyperX} (Hamming graph)
(HX)~\cite{ahn2009hyperx}. Other proposed topologies in this family include 
Flexfly~\cite{wen2016flexfly},
Galaxyfly~\cite{lei2016galaxyfly}, Megafly~\cite{flajslik2018megafly},
projective topologies~\cite{camarero2016projective}, HHS~\cite{azizi2016hhs},
and others~\cite{rahman2018load, kathareios2015cost, penaranda2016k}. 
All these networks have different structure and thus different potential for
multipathing~\cite{besta2019fatpaths}; in Figure~\ref{fig:topologies}, we
illustrate example paths between a pair of routers. Importantly, in most of
these networks, \emph{unlike in fat trees, different paths between two endpoints
usually have different lengths}~\cite{besta2019fatpaths}.

Finally, many routing designs can be used with any topology, including
\textbf{traditional ones} such as meshes.

\subsection{Routing Concepts and Related}
\label{sec:concepts}


We often refer to three interrelated sub-problems for routing: 
\includegraphics[scale=0.2,trim=0 16 0 0]{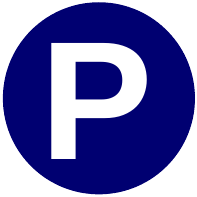} \textbf{Path selection},
\includegraphics[scale=0.2,trim=0 16 0 0]{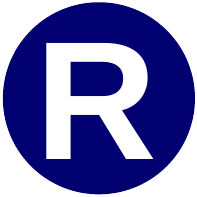} \textbf{Routing} itself,
and \includegraphics[scale=0.2,trim=0 16 0 0]{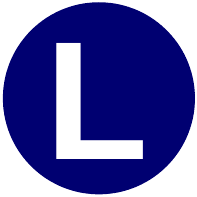} \textbf{Load
balancing}.
Path selection~\includegraphics[scale=0.2,trim=0 16 0 0]{P.pdf} determines
\emph{which paths} can be used for sending a given packet.  Routing
itself~\includegraphics[scale=0.2,trim=0 16 0 0]{R.pdf} answers a question on
\emph{how} the packet finds a way to its destination. Load
balancing~\includegraphics[scale=0.2,trim=0 16 0 0]{L.pdf} determines
\emph{which path (out of identified alternatives)} should be used for sending a
packet \emph{to maximize performance} and \emph{minimize congestion}. 
%

\subsection{Routing Schemes}

We consider routing schemes (designs) that can be loosely grouped into specific \textbf{protocols}
(e.g., OSPF~\cite{moy1997ospf}), \textbf{architectures} (e.g.,
PortLand~\cite{niranjan2009portland}), and general \textbf{strategies and
techniques} (e.g., ECMP~\cite{hopps2000analysis} or spanning
trees~\cite{perlman1985algorithm}).
Overall, a protocol or a strategy often addresses a \emph{specific}
networking problem, rarely more than one.  Contrarily, a routing architecture
usually delivers a \emph{complete routing solution} and it often addresses more
than one, and often all, of the above-described problems.
All these designs are almost always developed in the context of a specific
\textbf{network stack}, also referred to as \textbf{network architecture}, that
we describe next.

\subsection{Network Stacks}
\label{sec:back_arch}

We focus on data centers and high-performance systems. Thus, we target 
Ethernet \&
{TCP/IP}, and traditional HPC networks ({InfiniBand},
{Myrinet}, OmniPath, and others). 
%

\subsubsection{Ethernet \& TCP/IP}
\label{sec:back_eth}

In the TCP/IP protocol stack, two layers of addressing are used. On
\textbf{Layer~2 (L2)}, \emph{Ethernet} (\emph{MAC}) addresses are used to
uniquely identify endpoints, while on \textbf{Layer~3 (L3)}, \emph{IP}
addresses are assigned to endpoints. Historically, the Ethernet layer is not
supposed to be routable: MAC addresses are only used within a bus-like topology
where no routing is required. In contrast, the IP layer is designed to be
routable, with a hierarchical structure that allows scalable routing over a
worldwide network (the \emph{Internet}). More recently, vendors started to
provide routing abilities on the Ethernet layer for pragmatic reasons: since
the Ethernet layer is effectively transparent to the software running on the
endpoints, such solutions are easy to deploy.  Additionally, the Ethernet
interconnect of a cluster can usually be considered homogeneous, while the IP
layer is used to route \emph{between} networks and needs to be highly
interoperable. 

Since Ethernet was not designed to be routable, there are several restrictions
on routing protocols for Ethernet: First, the network \emph{cannot} modify any
fields in the packets (control-data plane separation is key in self-configuring
Ethernet devices). There is no mechanism like the \emph{TTL} field in the IP
header that allows the network to detect cyclic routing. Second, Ethernet
devices come with pre-configured, effectively random addresses. This implies
that there is no structure in the addresses that would allow for a scalable
routing implementation: Each switch needs to keep a lookup table with entries
for each endpoint in the network. Third, since the network is expected to
  self-configure, Ethernet routing schemes must be robust to the addition and
  removal of links. These restrictions shape many routing schemes for Ethernet:
  Spanning trees are commonly used to guarantee loop-freedom under any
  circumstances, and more advanced schemes often rely on wrapping Ethernet
  frames into a format more suitable for routing at the edge
  switches~\cite{niranjan2009portland}.

Another intricacy of the TCP/IP stack is that flow control is only implemented
in \textbf{Layer~4 (L4)}, the transport layer. This means that the network is
not supposed to be aware of and responsible for load balancing and resource
sharing; rather, it should deliver packets to the destination on a best-effort
basis. In practice, most advanced routing schemes violate this separation and
are \emph{aware} of TCP flows, even though flow control is still left to the
endpoint software~\cite{handley2017re}. Many practical problems are caused by
the interaction of TCP flow control with decisions in the routing layer, and
such problems are often discussed together with routing schemes, even though
they are completely independent of the network topology (e.g., the \emph{TCP
incast problem}).

\ifconf
\fi

{Traditional Ethernet is \emph{lossy}: when packet buffers are full, packets
are dropped. Priority Flow Control (PFC)~\mbox{\cite{decusatis2013handbook}}
addresses this by allowing a switch to notify another (upstream) switch
\emph{with special ``pause'' frames} to stop sending frames until further
notice, if the buffer occupancy in the first switch is above a certain
threshold.}
{Another extension of Ethernet towards technologies traditionally associated
with HPC is the incorporation of Remote Direct Memory Access (RDMA) using the
RDMA over Converged Ethernet (RoCE)~\mbox{\cite{infiniband2014rocev2}}
protocol, which enriches the Ethernet with the RDMA communication
semantics.}

\ifconf
\fi

\subsubsection{InfiniBand}
\label{sec:back_ib}

The InfiniBand (IB) architecture is a switched fabric design and is intended
for high-performance and system area network (SAN) deployment scales. Up to
  49,151 endpoints (physical or virtual), addressed by a \unit[16]{bit}
  \emph{local identifier} (LID), can be arranged in a so called
  \textbf{subnet}, while the remaining address space is reserved for multicast
  operations within a subnet.  Similar to the modern datacenter Ethernet (L2)
  solutions, these IB subnets are routable to a limited extent with switches
  supporting unicast and multicast forwarding tables, flow control, and other
  features which do not require modification of in-flight packet headers.
  Theoretically, multiple subnets can be connected by IB routers --- performing
  the address translation between the subnets --- to create larger SANs
  (effectively \textbf{L3} domains), but this impedes performance due to the
  additionally required \emph{global routing header} (GRH) and is
  rarely used in practice.

IB natively supports RDMA and
atomic operations. The necessary (for high performance) lossless
packet forwarding within IB subnets is realized through link-level,
credit-based flow control~\cite{Dally:2003:PPI:995703}.
Software-based and latency impeding solutions to
achieve reliable transmissions, as for example in TCP, are therefore not
required.  While switches have the capability to drop deadlocked packets that
reside for extended time periods in their buffers, they cannot identify
livelocks, such as looping unicast or multicast packets induced by cyclic
routing.  Hence, the correct and acyclic routing configuration is offloaded to
a centralized controller, called \emph{subnet manager}, which configures
connected IB devices, calculates the forwarding tables with implemented
topology-agnostic or topology-aware routing algorithms, and monitors the
network for failures.  Therefore, most routing algorithms either focus on
\emph{minimal path length} to guarantee loop-freedom, or are derivatives of the
Up*/Down* routing protocol~\cite{lysne_load_2001, flich_improving_2002} which
can be viewed as a generalization of the spanning tree protocol of Ethernet
networks. Besides this oblivious, \emph{destination-based} routing approach, IB
also supports \emph{source-based} routing, but unfortunately only for a limited
traffic class reserved for certain management packets.

The subnet manager can configure the InfiniBand network with a few flow control
features, such as quality-of-service to prioritize traffic classes over others
or congestion control mechanism to throttle ingest traffic. However, adhering
to the correct service levels or actually throttling the packet generation is
left to the discretion of the endpoints. Similarly, in sacrifice for lowest
latency and highest bandwidth, IB switches have limited support for common
capabilities found in Ethernet, for example VLANs, firewalling, or other
security-relevant functionality. Consequently, some of these have been
implemented in software at the endpoints on top of the IB transport protocol,
e.g., TCP/IP via IPoIB, whenever the HPC community deemed it necessary.

\subsubsection{Other HPC Network Designs}

%
%
Cray's \textbf{Aries}~\cite{aries} is a packet-switched interconnect designed for
high performance and deployed on the Cray XC systems. Aries
adopts a dragonfly topology, where nodes within groups are interconnected
with a two-dimensional all-to-all structure (i.e., routers in one dragonfly group effectively form
a flattened butterfly, cf.~Figure~\ref{fig:topologies}). 
Being designed for high-performance systems, it allows nodes to communicate
with RDMA operations (i.e., put, get, and atomic operations). The routing is
destination-based, and the network addresses are tuples composed by a node
identifier (18-bit, max 262,144 nodes), the memory domain handle (12-bit) that
identifies a memory segment in the remote node, and an offset (40-bit) within
this segment. The Aries switches employ wormhole
routing~\cite{dally2004principles} to minimize the per-switch required
resources.
Aries does not support VLANs or QoS mechanisms, and its stack design does not
match that of Ethernet. Thus, we define the (software) layer at which the
Aries routing operates as \textbf{proprietary}.

\textbf{Slingshot}~\cite{slingshot} is the next-generation Cray network.
It implements a DF topology with fully-connected groups. Slingshot
can switch two types of traffic: 
RoCE (using \textbf{L3}) and
proprietary. Being able to manage RoCE traffic, a Slingshot system can be
interfaced directly to data centers, while the proprietary traffic (similar to
Aries, i.e., RDMA-based and small-packet headers) can be generated from within the
system, preserving high performance. 
Cray Slingshot supports VLANs, QoS, and endpoint congestion
mitigation. 

IBM's \textbf{PERCS}~\cite{ibm-percs-network} is a two-level direct
interconnection network designed to achieve high bisection bandwidth and avoid
external switches. Groups of 32 compute nodes (made of four IBM POWER7 chips)
are fully connected and organized in supernodes. Each supernode has 512 links
connecting it to other supernodes.  Depending on the system size (max 512
supernodes), each supernode pair can be connected with one or multiple links.
\iftr
The PERCS Hub Chip connects the POWER7 chips within a compute node between
themselves and with the rest of the network. The Hub Chip participates to the
cache coherence protocol and is able to fetch/inject data directly from the
processors' L3 caches.
\fi
PERCS supports RDMA, hardware-accelerated collective operations, direct-cache
(L3) network access, and enables applications to switch between different
routing modes. Similarly to Aries, PERCS routing operates on a
\emph{proprietary} stack.

We also summarize other HPC oriented proprietary interconnects. Some of them
are no longer manufactured; we include them for the completeness of our
discussion of path diversity.
Myricom's \textbf{Myrinet}~\cite{boden1995myrinet} is a local area
\emph{massively parallel processor} network, designed to connect thousands of
small compute nodes. A more recent development, \textbf{Myrinet Express
(MX)}~\cite{geoffray2004myrinet}, provides more functionalities in its network
interface cards (NICs). \textbf{Open-MX}~\cite{goglin2008design} is a
communication layer that offers the MX API on top of the Ethernet hardware.
Quadrics' \textbf{QsNet}~\cite{petrini2003performance, petrini2002quadrics}
integrates local memories of compute nodes into a single \emph{global virtual
address space}.
{Moreover, Intel introduced
\textbf{OmniPath}~\mbox{\cite{birrittella2015intel}}, an architecture for a
tight integration of CPU, memory, and storage units.}
{Other HPC interconnects are Atos' Bull eXascale Interconnect
(BXI)~\mbox{\cite{derradji2015bxi}} and EXTOLL's
interconnect~\mbox{\cite{neuwirth2015scalable}}.}
Many of these architectures feature some form of programmable
NICs~\cite{boden1995myrinet, petrini2002quadrics}.
{Finally, there exist routing protocols for specific
low-diameter topologies, for example for SF~\mbox{\cite{yebenes2017improving}}
or DF~\mbox{\cite{maglione2017scalable}}. However, they usually do
not support multipathing or non-minimal routing.}

\ifconf
\fi

\subsection{Focus of This Work}

In our investigation, \emph{we focus on \textbf{routing}}. Thus, in the
Ethernet and TCP/IP landscape, we focus on designs associated with Layer~2 (L2,
Data Link Layer) and Layer~3 (L3, Internet Layer), cf.~\cref{sec:back_eth}. As
most of congestion control and load balancing are related to
higher layers, we only describe such schemes whenever they are parts of the
associated L2 or L3 designs.
In the InfiniBand landscape, we focus on the subnet and L3 related schemes,
cf.~\cref{sec:back_ib}.

\begin{table*}[hbtp]
%
\setlength{\tabcolsep}{1pt}
\ifsq\renewcommand{\arraystretch}{0.8}\fi
\centering
\scriptsize
\sf
\begin{tabular}{llllllllll}
%
%
\toprule
\multirow{2}{*}{ \makecell[c]{\textbf{Routing Scheme}\\\textbf{(Name, Abbreviation, Reference)}}} & \multirow{2}{*}{ \makecell[c]{\textbf{Related}\\\textbf{concepts}\\(\cref{sec:concepts})}} & \multirow{2}{*}{\makecell[c]{\textbf{Stack}\\\textbf{Layer}\\(\cref{sec:back_arch})}} & \multicolumn{6}{c}{Features of schemes} & \multirow{2}{*}{\makecell[c]{\textbf{Additional remarks and clarifications}}} \\
\cmidrule(lr){4-9}
 & & & \textbf{SP} & \textbf{NP} & \textbf{MP} & \textbf{DP} & \textbf{ALB} & \textbf{AT} & 
%
%
%
\vspace{0.5em} \\
\midrule
\multicolumn{10}{c}{\textbf{General routing building blocks (classes of routing schemes)}} \\
\midrule
Simple Destination-based routing & \includegraphics[scale=0.15,trim=0 16 0 0]{R.pdf} & L2, L3 & \faThumbsOUp & \faThumbsUp$^*$ & \faTimes & \faTimes & \faTimes & \faThumbsOUp & $^*$Care must be taken not to cause cyclic dependencies \\
Simple Source-based routing (SR) & \includegraphics[scale=0.15,trim=0 16 0 0]{R.pdf} & L2, L3 & \faThumbsOUp & \faThumbsOUp & \faThumbsUp$^*$ & \faThumbsUp$^*$ & \faTimes & \faThumbsOUp  & \parbox[t]{8cm}{Source routing is difficult to deploy in practice, but it is more flexible than\\ destination-based routing. $^*$As endpoints know the physical topology,\\ multipathing should be easier to realize than in destination routing.} \\
Simple Minimal routing & \includegraphics[scale=0.15,trim=0 16 0 0]{P.pdf} & L2, L3 &  \faThumbsOUp & \faTimes & \faTimes & \faTimes & \faTimes & \faThumbsOUp  & Easy to deploy, numerous designs fall in this category \\
\midrule
\multicolumn{10}{c}{\textbf{Specific routing building blocks (concrete protocols or concrete protocol families)}} \\
\midrule
Equal-Cost Multipathing (ECMP)~\cite{hopps2000analysis} & \includegraphics[scale=0.15,trim=0 16 0 0]{R.pdf} \includegraphics[scale=0.15,trim=0 16 0 0]{L.pdf} & L3 & \faThumbsOUp & \faTimes & \faThumbsOUp & \faTimes & \faTimes & \faThumbsOUp  & In ECMP, all routing decisions are local to each switch. \\ 
\makecell[l]{Spanning Trees (ST)~\cite{perlman1985algorithm}} & \includegraphics[scale=0.15,trim=0 16 0 0]{P.pdf} & L2 & \faThumbsUp$^{*}$ & \faThumbsUp$^*$ & \faTimes & \faTimes & \faTimes & \faThumbsOUp  & $^*$\makecell[l]{The ST protocol offers shortest paths but only within one spanning tree.} \\ 
Packet Spraying (PR)~\cite{dixit2013impact} & \includegraphics[scale=0.15,trim=0 16 0 0]{L.pdf} & L2, L3 & \faThumbsOUp & \faTimes & \faThumbsOUp & \faTimes & \faTimes & \faThumbsOUp  & \makecell[l]{One selects output ports with round-robin~\cite{dixit2013impact} or randomization~\cite{sen2013localflow}.} \\
%
%
Virtual LANs (VLANs) & \includegraphics[scale=0.15,trim=0 16 0 0]{P.pdf} & L2 & \faThumbsUp$^*$ & \faThumbsUp$^*$ & \faTimes & \faTimes & \faTimes & \faThumbsOUp & \parbox[t]{8cm}{$^*$VLANs by itself does not focus on multipathing, and it inherits\\ spanning tree limitations, but it is a key part of multipathing architectures.} \\
\makecell[l]{IP Routing Protocols} & \includegraphics[scale=0.15,trim=0 16 0 0]{R.pdf} & L2, L3 & \faThumbsOUp & \faTimes & \faTimes & \faTimes & \faTimes & \faThumbsOUp  & Examples are OSPF~\cite{moy1997ospf}, IS-IS~\cite{oran1990osi}, EIGRP~\cite{pepelnjak1999eigrp}. \\
\makecell[l]{Location--Identification Separation (LIS)} & \includegraphics[scale=0.15,trim=0 16 0 0]{R.pdf} & L2, L3 & \faThumbsOUp$^*$ & \faThumbsOUp$^*$ & \faTimes$^*$ & \faTimes$^*$ & \faTimes & \faThumbsOUp & \parbox[t]{8cm}{$^*$LIS by itself does not focus on multipathing and path diversity, \\but it may facilitate developing a multipathing architecture.} \\
Valiant load balancing (VLB)~\cite{valiant1982scheme} & \includegraphics[scale=0.15,trim=0 16 0 0]{R.pdf} \includegraphics[scale=0.15,trim=0 16 0 0]{P.pdf} \includegraphics[scale=0.15,trim=0 16 0 0]{L.pdf} & L2, L3 & \faTimes & \faThumbsOUp & \faTimes & \faTimes & \faTimes & \faThumbsOUp  & --- \\
UGAL~\cite{kim2008technology} & \includegraphics[scale=0.15,trim=0 16 0 0]{R.pdf} \includegraphics[scale=0.15,trim=0 16 0 0]{P.pdf} \includegraphics[scale=0.15,trim=0 16 0 0]{L.pdf} & L2, L3 & \faThumbsOUp & \faThumbsOUp & \faThumbsUp$^*$ & \faTimes & \faThumbsOUp & \faThumbsOUp  & \makecell[l]{UGAL means Universal Globally-Adaptive Load balanced routing.} \\ 
Network Address Aliasing (NAA) & \includegraphics[scale=0.15,trim=0 16 0 0]{L.pdf} & \makecell[l]{L3, subnet} & \faThumbsOUp$^{*}$ & \faThumbsOUp$^{*}$ & \faThumbsOUp$^{*}$ & \faThumbsOUp$^{*}$ & \faThumbsOUp$^{*}$ & \faThumbsOUp  & \parbox[t]{8cm}{NAA is based on IP aliasing in Ethernet networks~\cite{rfc2835} and\\virtual ports via LID mask control (LMC) in InfiniBand~\cite[Sec.~7.11.1]{infiniband_trade_association_infinibandtm_2015}.\\$^*$Depending on how a derived scheme implements it.} \\
Multi-Railing & \includegraphics[scale=0.15,trim=0 16 0 0]{P.pdf} & \makecell[l]{L2, L3, subn.} & \faThumbsOUp$^{*}$ & \faThumbsOUp$^{*}$ & \faThumbsOUp & \faThumbsOUp$^{*}$ & \faThumbsOUp$^{*}$ & \faThumbsOUp & \makecell[l]{$^*$Depending on how a derived scheme implements it.} \\
Multi-Planes & \includegraphics[scale=0.15,trim=0 16 0 0]{P.pdf} & \makecell[l]{L2, L3, subn.} & \faThumbsOUp$^{*}$ & \faThumbsOUp$^{*}$ & \faThumbsOUp & \faThumbsOUp & \faThumbsOUp$^{*}$ & \faThumbsOUp & \makecell[l]{$^*$Depending on how a derived scheme implements it.} \\
\bottomrule
\end{tabular}
\vspace{-1em}
\caption{
\textmd{
\textbf{Comparison of simple routing building blocks (often used as parts of
more complex routing schemes in Table~\ref{tab:intro}). Rows are sorted
chronologically. We focus on how well the compared schemes utilize path
diversity.} 
\textbf{``Related concepts''} indicates the associated routing concepts
described in~\cref{sec:concepts}. 
\textbf{``Stack Layer''} indicates the location of each routing scheme in the
TCP/IP or InfiniBand stack (cf.~\cref{sec:back_arch}).
\textbf{SP}, \textbf{NP}, \textbf{MP}, \textbf{DP}, \textbf{ALB}, and
\textbf{AT} illustrate whether a given routing scheme supports various aspects
of path diversity. Specifically:
\textbf{SP}: A given scheme enables using arbitrary \textbf{shortest} paths.
\textbf{NP}: A given scheme enables using arbitrary \textbf{non-minimal} paths.
\textbf{MP}: A given scheme enables \textbf{multipathing} (between two hosts).
\textbf{DP}: A given scheme considers \textbf{disjoint} paths.
\textbf{ALB}: A given scheme offers \textbf{adaptive load balancing}.
\textbf{AT}: A given scheme works with an \textbf{arbitrary topology}.
%
%
%
\faThumbsOUp: A given scheme does offer a given feature. \faThumbsUp: A given
scheme offers a given feature in a limited way. \faTimes: A given scheme
does not offer a given feature. $^*$Explanations in remarks.
}
}
\ifsq\vspace{-2em}\fi
\label{tab:blocks}
\end{table*}

\section{Taxonomy of Routing Schemes}
\label{sec:taxonomy}

We first identify criteria for categorizing the considered routing designs. We
focus on how well these designs utilize path diversity. These criteria
are used in Tables~\ref{tab:blocks}--\ref{tab:intro}.
%
%
Specifically, we analyze whether a given scheme enables using (1) arbitrary
\textbf{shortest} paths and (2) arbitrary \textbf{non-minimal} paths.
Moreover, we consider whether a studied scheme enables (3)
\textbf{multipathing} (between two hosts) and whether these paths can be (4)
\textbf{disjoint}. Finally, we investigate (5) the support for \textbf{adaptive
load balancing} across exposed paths between router pairs and (6) compatibility
with an \textbf{arbitrary topology}.
In addition, we also indicate the \textbf{location} of each routing scheme in
the networking stack\footnote{\scriptsize We consider protocols in both Data Link (L2)
and Network (L3) layers. However, we abstract away hardware details and use a
term ``router'' for both L2 switches and L3 routers, \emph{unless describing
a specific \ul{switching} protocol (to avoid confusion)}.}.
We also indicate whether a given multipathing scheme focuses on
\textbf{performance} or \textbf{resilience} (i.e., to provide
backup paths in the event of failures).
Next, we identify whether supported paths come \textbf{with certain
restrictions}, e.g., whether they are offered only within a spanning tree.
Finally, we also broadly categorize the analyzed routing schemes into
\textbf{basic} and \textbf{complex} ones. The former are usually specific
protocols or classes of protocols, used as
building blocks of the latter.

%

\section{Simple Routing Building Blocks}

We now present simple routing schemes, summarized in Table~\ref{tab:blocks}, that are
usually used as building blocks for more complex routing designs.
For each described scheme, we indicate \emph{what aspects of routing (as
described in~\cref{sec:concepts}) this scheme focuses on:
\includegraphics[scale=0.2,trim=0 16 0 0]{P.pdf} path selection,
\includegraphics[scale=0.2,trim=0 16 0 0]{R.pdf} routing itself,
\includegraphics[scale=0.2,trim=0 16 0 0]{L.pdf} or load balancing}.
We consider both \emph{general} classes of schemes (e.g., overall
destination-based routing) and also \emph{specific} protocols (e.g., Valiant
routing~\cite{valiant1982scheme}).

Note that, in addition to schemes focusing on multipathing, we also describe
designs that do \emph{not} explicitly enable it. This is because these designs
are often used as key building blocks of architectures that provide
multipathing. An example is a simple spanning tree mechanism, that -- on its
own -- does not enable any form of multipathing, but is a basis of numerous
designs that enable it~\cite{ieee2001mstp, stephens2012past}.

\subsection{Destination-Based Routing Protocols}

The most common approach to \includegraphics[scale=0.2,trim=0 16 0 0]{R.pdf}
\emph{routing} are destination-based routing schemes. Each router holds a
\emph{routing table} that maps any destination address to a next-hop output
port. No information apart from the destination address is used, and the packet
does not need to be modified in transit. 
In this setup, it is important to differentiate the physical network topology
(typically modeled as an undirected graph, since all practically used network
technologies use full-duplex links, cf.~\cref{sec:nettopo}) from the
\emph{routing graph}, which is naturally \emph{directed} in destination-based
schemes. In the routing graph, there is an edge from node $a$ to node $b$ iff
there is a routing table entry at $a$ indicating $b$ as the next hop
destination. 

\iftr
\iftrcol\textcolor{blue}{\fi
Typically, the lookup table is implemented using longest-prefix matching, which
allows entries with an identical address prefix and identical output port to be
compressed into one table slot. This method is especially well suited to
hierarchically organized networks. In general, longest-prefix matching is not
required: it is feasible and common to keep uncompressed routing tables, e.g.,
in Ethernet routing.
\iftrcol}\fi
\fi

Simple destination-based routing protocols can only provide a single path
between any source and destination, \emph{but this path can be non-minimal}.
For non-minimal paths, special care must be taken to not cause cyclic
routing: this can happen when the routing tables of different routers are not
consistent, cf.~\emph{property preserving network updates}~\cite{domke_scheduling-aware_2016}.
In a configuration without routing cycles, the routing graph for a
fixed destination node is a tree rooted in the destination.

\subsection{Source Routing (SR)}

Another \includegraphics[scale=0.2,trim=0 16 0 0]{R.pdf} \emph{routing}
scheme is \emph{source routing} (SR). Here, the route from source to destination is
computed \emph{at the source}, and then attached to the packet before it is
injected into the network. Each switch then reads (and possibly removes) the next hop
entry from the route, and forwards the packet there. Compared to destination
based routing, this allows for far more flexible path
selection~\cite{jyothi2015towards}. Yet, now the endpoints need to be
aware of the network topology to make viable routing choices. 

Source routing is rarely deployed in practice.  Still, it could enable superior
routing decisions (compared to destination based routing) in terms of utilizing
path diversity, as endpoints know the physical topology. There are recent
proposals on how to deploy source routing in practice, for example with the
help of OpenFlow~\cite{jyothi2015towards}, or with packet encapsulation
(IP-in-IP or MAC-in-MAC)~\cite{greenberg2008towards, mac-in-mac,
greenberg2009vl2}.
Source routing can also be achieved to some degree with Multiprotocol
Label Switching (MPLS)~\cite{rosen2001multiprotocol}, a technique
in which a router forwards packets based on \emph{path labels} instead of
\emph{network addresses} (i.e., the MPLS label assigned to a packet
can represent a path to be chosen~\cite{rosen2001multiprotocol, xu2017unified}).

\subsection{Minimal Routing Protocols}

A common approach to \includegraphics[scale=0.2,trim=0 16 0 0]{P.pdf}
\emph{path selection} is to only use \emph{minimal} paths: Paths that are no
longer than the shortest path between their endpoints.  Minimal paths are
preferrable for routing because they minimize network resources
consumed for a given volume of traffic, which is crucial to achieve good
performance at high load. 

An additional advantage of minimal paths is that they guarantee loop-free
routing in destination-based routing schemes. For a known, fixed topology, the
routing tables can be configured to always send packets along shortest paths.
Since every hop along any shortest path will decrease the shortest-path
distance to the destination by one, the packet always reaches its destination
in a finite number of steps.

\iftr
\iftrcol\textcolor{blue}{\fi
To construct shortest-path routing tables, a variation of the
Floyd-Warshall all-pairs shortest path algorithm~\cite{floyd1962algorithm} can
be used. Here, besides the shortest-path distance for all router pairs, one
also records the out-edge at a given router (i.e., the output port) for the
\emph{first} step of a shortest path to any other router. 
Other schemes are also applicable, for example an algorithm by Suurballe
and Tarjan for finding shortest pairs of edge-disjoint
paths~\cite{suurballe1984quick}.
\iftrcol}\fi
\fi

Basic minimal routing does not consider multipathing. However, schemes such as
Equal-Cost Multipathing (ECMP) extend minimal routing to multipathing
(\cref{sec:ecmp}).

\subsection{Equal-Cost Multipathing (ECMP)}
\label{sec:ecmp}

\emph{Equal-Cost Multipathing}~\cite{hopps2000analysis} routing is an extension
of simple destination-based \includegraphics[scale=0.2,trim=0 16 0 0]{R.pdf}
routing that specifically exploits the properties of minimal paths. Instead of
having only one entry per destination in the routing tables, multiple next-hop
options are stored. In practice, ECMP is used with minimal paths, because
using non-minimal ones may lead to routing loops.
Now, any router can make an arbitrary choice among these next-hop
options. The resulting routing will still be loop-free and only use minimal
paths. 

ECMP allows to use a greater variety of paths compared to simple
destination-based routing. Since now there may be multiple possible paths
between any pair of nodes, a mechanism for \includegraphics[scale=0.2,trim=0 16
0 0]{L.pdf} \emph{load balancing} is needed. Typically, ECMP is used with a
simple, oblivious scheme similar to packet spraying (\cref{sec:ps}), but on a
per-flow level to prevent packet reordering~\cite{chim2004traffic}: each switch
chooses a pseudo-random next hop port among the shortest paths based on a hash
computed from the flow parameters, aiming to obtain an even distribution of
load over all minimal paths (some variations of such simple per-flow scheme
were proposed, for example Table-based Hashing~\cite{sridharan2005achieving} or
FastSwitching~\cite{zinin2002routing}). Yet, random assignments \emph{do not}
imply uniform load balancing in general, and more advanced schemes such as
\emph{Weighted Cost Multipathing} (WCMP)~\cite{zhang2012optimizing,
zhou2014wcmp} aim to improve this.
In addition, ECMP natively does not support adaptive load balancing.  This is
addressed by many network architectures described in
Section~\ref{sec:architectures} and by direct extensions of ECMP, such as
\emph{Congestion-Triggered Multipathing} (CTMP)~\cite{sohn2006congestion} or
Table-based Hashing with Reassignments (THR)~\cite{chim2004traffic}.

\subsection{Spanning Trees (ST)}

Another approach to \includegraphics[scale=0.2,trim=0 16 0 0]{P.pdf} \emph{path
selection} is to restrict the topology to a spanning tree.  Then, the routing
graph becomes a tree of bi-directional edges which guarantees the absence of
cycles as long as no router forwards packets back on the link that the packet
arrived on. This can be easily enforced by each router without any global
coordination. Spanning tree based solutions are popular for auto-configuring
protocols on changing topologies.  However, simple spanning tree-based routing
can leave some links completely unused if the network topology is not a tree.
Moreover, shortest paths within a spanning tree are not necessarily shortest
when considering the whole topology.  Spanning tree based solutions are an
alternative to minimal routing to ensure loop-free routing in destination-based
routing systems. They allow for non-minimal paths at the cost of not using
network resources efficiently and have been used as a building block in schemes
like SPAIN~\cite{mudigonda2010spain}.
A single spanning tree does not enable multipathing between two endpoints.
However, as we discuss in Section~\ref{sec:architectures}, different network
architectures use spanning trees to enable
multipathing~\cite{stephens2012past}.

\subsection{Packet Spraying}
\label{sec:ps}

A fundamental concept for \includegraphics[scale=0.2,trim=0 16 0 0]{L.pdf}
\emph{load balancing} is per-packet load balancing. In the basic variant,
\emph{random packet spraying}~\cite{dixit2013impact}, each packet
is sent over a randomly chosen path selected from a (static) set of possible
paths.
The key difference from ECMP is that modern ECMP spreads flows, not packets.
Typically, packet spraying is applied to multistage networks, where many 
equal length paths are available and a random path among these can be chosen
by selecting a random upstream port at each router.
Thus, simple packet spraying natively considers, enables, and uses multipathing.

In TCP/IP architectures,
per-packet load balancing is often not considered due to the negative effects
of packet reordering on TCP flow control; but these effects can still be
reduced in various ways~\cite{dixit2013impact, handley2017re},
for example by spraying not single packets but series of packets,
  such as flowlets~\cite{vanini2017letflow} or flowcells~\cite{he2015presto}.
Moreover, basic random packet spraying is an \emph{oblivious} load balancing method, as it does
not use any information about network congestion. However, in some topologies,
for example in fat trees, it can still guarantee optimal performance as long as
  it is used for all flows. Unfortunately, this is no longer true as soon as the topology
  looses its symmetry due to link failures~\cite{zhou2014wcmp}.

\subsection{Virtual LANs (VLANs)}

Virtual LANs (VLANs)~\cite{ls2006ieee} were originally used for isolating
Ethernet broadcast domains. They have recently been used to
implement multipathing.
Specifically, once a VLAN is assigned to a given spanning tree, changing the
VLAN tag in a frame results in sending this frame over a different path,
associated with a different spanning tree (imposed on the same physical
topology). Thus, VLANs -- in the context of multipathing -- primarily address
path selection \includegraphics[scale=0.2,trim=0 16 0 0]{P.pdf}.

\subsection{Simple IP Routing}

We explicitly distinguish a class of established IP routing protocols
\includegraphics[scale=0.2,trim=0 16 0 0]{R.pdf}, such as
OSPF~\cite{moy1997ospf}
or IS-IS~\cite{oran1990osi}.  They are often used as
parts of network architectures. Despite being generic (i.e., they can be used
with any topology), they do not natively support multipathing.

\subsection{Location--Identification Separation (LIS)}

In Location--Identification Separation (LIS), used in some architectures, a
routing scheme \includegraphics[scale=0.2,trim=0 16 0 0]{R.pdf} \emph{separates
the physical location of a given endpoint from its logical identifier}. In this
approach, the logical identifier of a given endpoint (e.g., its IP address used
in an application) does not necessarily indicate the physical location of this
endpoint in the network. A mapping between identifiers and addresses can be
stored in a distributed hashtable (DHT) maintained by
switches~\cite{kim2008floodless} or hosts, or it can be provided by a directory
service (e.g., using DNS)~\cite{greenberg2009vl2}.  This approach enables more
scalable routing~\cite{farinacci2013locator}.  Importantly, it may facilitate
multipathing by -- for example -- maintaining multiple \emph{virtual}
topologies defined by different mappings in DHTs~\cite{jain2011viro}.

\subsection{Valiant Load Balancing (VLB)}

To facilitate non-minimal \includegraphics[scale=0.2,trim=0 16 0 0]{R.pdf}
\emph{routing}, additional information apart from the destination address can
be incorporated into a destination-based routing protocol. An established and
common approach is \emph{Valiant routing}~\cite{valiant1982scheme}, where this
additional information is an arbitrary intermediate router~$R$ that can be selected
at the source endpoint. The routing is
divided into two parts: first, the packet is minimally routed to $R$; 
then, it is minimally routed to the actual
destination. VLB has aspects of source routing, namely the choice
of $R$ and the modification of the packet in flight, while
most of the routing work is done in a destination-based way. As such, VLB
natively does not consider multipathing.
VLB also incorporates a specific \includegraphics[scale=0.2,trim=0
16 0 0]{P.pdf} \emph{path selection} (by
selecting the intermediate node randomly).  This also provides simple, oblivious
\includegraphics[scale=0.2,trim=0 16 0 0]{L.pdf} \emph{load balancing}. 

\subsection{\hspace{-0.5em}\mbox{Universal Globally-Adaptive Load Balanced (UGAL)}}

Universal Globally-Adaptive Load balanced (UGAL)~\cite{kim2008technology} is an
extension of VLB that enables more advantageous routing decisions
\includegraphics[scale=0.2,trim=0 16 0 0]{R.pdf}
\includegraphics[scale=0.2,trim=0 16 0 0]{P.pdf} in the context of load
balancing \includegraphics[scale=0.2,trim=0 16 0 0]{L.pdf}.
Specifically, when a packet is to be routed, UGAL either selects a path
determined by VLB, or a minimum one. The decision usually depends on the
congestion in the network.  Consequently, UGAL considers multipathing in its
design: consecutive packets may be routed using different paths.

\subsection{Network Address Aliasing (NAA)}

Network Address Aliasing~(NAA) 
is a building block to support multipathing, especially in InfiniBand-based
networks. Network Address Aliasing, also known as IP aliasing in Ethernet
networks~\cite{rfc2835} or port virtualization via LID mask control (LMC) in
InfiniBand~\cite[Sec.~7.11.1]{infiniband_trade_association_infinibandtm_2015},
is a technique that \emph{assigns \textbf{multiple} identifiers to \textbf{the
same} network endpoint}.  This allows the routing protocols to increase the
path diversity between two endpoints, and it was used both as a fail-over
(enhancing resilience)~\cite{vishnu_automatic_2007} or for \includegraphics[scale=0.2,trim=0 16 0
0]{L.pdf} load balancing the traffic (enhancing performance)~\cite{domke_hyperx_2019}. In
particular, due to the destination-based routing --- where a path is only
defined by the given destination address; as mandated by the InfiniBand
standard~\cite{infiniband_trade_association_infinibandtm_2015} --- this address
aliasing is the only standard-conform and software-based solution to enable
multiple disjoint paths between an IB source and a destination port.


\subsection{Multi-Railing and Multi-Planes}
Various HPC systems employ \emph{\textbf{multi-railing}}: using multiple
injection ports per node into a single
topology~\cite{gsic_tokyo_institute_of_technology_tsubame3.0_2017,
wolfe_preliminary_2017}.
Another common scheme is \emph{\textbf{multi-plane}} topologies, where nodes
are connected to a set of disjoint topologies, either
similar~\cite{gsic_tokyo_institute_of_technology_tsubame2.5_2013} or
different~\cite{matsuoka_a64fx_2019}. This is used to increase path diversity
and available throughput. However, this increased level of complexity also comes
with additional challenges for the routing protocols to utilize the hardware
efficiently.

\begin{table*}[hbtp]
\vspace{-0.5em}
\setlength{\tabcolsep}{1pt}
\renewcommand{\arraystretch}{0.6}
\centering
\scriptsize
\sf
\begin{tabular}{llllllllll}
%
%
\toprule
\multirow{2}{*}{ \makecell[c]{\textbf{Routing Scheme}}} & \multirow{2}{*}{\makecell[c]{\textbf{Stack}\\\textbf{Layer}}} & \multicolumn{6}{c}{Features of schemes} & \multirow{2}{*}{\makecell[c]{\textbf{Scheme}\\\textbf{used}}} & \multirow{2}{*}{\makecell[c]{\textbf{Additional remarks and clarifications}}} \\
\cmidrule(lr){3-8}
 & & \textbf{SP} & \textbf{NP} & \textbf{MP} & \textbf{DP} & \textbf{ALB} & \textbf{AT} & \\
%
%
%
\midrule
%
%
%
\multicolumn{10}{l}{\textbf{Related to Ethernet and TCP/IP (small clusters and general networks):}}\\
\midrule
OSPF-OMP (OMP)~\cite{villamizar1999ospf} & L3 & \faThumbsOUp & \faTimes & \faThumbsOUp & \faTimes & \faTimes & \faThumbsOUp & OSPF & \parbox[t]{8cm}{Cisco's enhancement of OSPF to the multipathing setting. Packets from\\ the same flow are forwarded using the same path.} \\ 

MPA~\cite{narvaez1999efficient} & L3 & \faThumbsUp$^*$ & \faThumbsUp$^*$ & \faThumbsUp$^*$ & \faTimes & \faTimes & \faThumbsOUp & --- & \makecell[l]{$^*$MPA only focuses on algorithms for \emph{generating} routing paths.} \\

SmartBridge~\cite{rodeheffer2000smartbridge} & L2 & \faThumbsOUp & \faTimes & \faTimes & \faTimes & \faTimes & \faThumbsOUp & ST & \parbox[t]{8cm}{SmartBridges improves ST; packets are sent between hosts\\using the shortest possible path in the network.} \\

MSTP~\cite{ieee2001mstp, de2006improving} & L2 & \faThumbsUp$^*$ & \faTimes$^*$ & \faThumbsOUp & \faTimes & \faTimes & \faThumbsOUp & ST+VLAN & \makecell[l]{$^*$Shortest paths are offered only within spanning trees.} \\ 

STAR~\cite{lui2002star} & L2 & \faThumbsOUp & \faTimes & \faTimes & \faTimes & \faTimes & \faThumbsOUp & ST & \parbox[t]{8cm}{STAR improves ST; frames are forwarded over alternate paths\\that are shorter than their corresponding ST path.} \\

LSOM~\cite{garcia2003lsom} & L2 & \faThumbsOUp & \faTimes & \faTimes & \faTimes & \faTimes & \faThumbsOUp & --- & \makecell[l]{LSOM supports mesh networks also in MAN. LSA manages state of links.}\\

AMP~\cite{gojmerac2003adaptive} & L3 & \faThumbsOUp & \faTimes & \faThumbsOUp & \faTimes & \faThumbsOUp & \faThumbsOUp & \makecell[l]{ECMP, OMP} & AMP extends ECMP and OSPF-OMP. \\

RBridges~\cite{perlman2004rbridges} & L2 & \faThumbsOUp & \faTimes & \faTimes & \faTimes & \faTimes & \faThumbsOUp & --- & --- \\

THR~\cite{chim2004traffic} & L3 & \faThumbsOUp & \faTimes & \faThumbsOUp & \faTimes & \faThumbsUp & \faThumbsOUp & ECMP & \parbox[t]{8cm}{Table-based Hashing with Reassignments (THR) extends ECMP, it\\ selectively reassigns some active flows based on load sharing statistics.} \\ 

GOE~\cite{iwata2004global} & L2 & \faThumbsUp$^*$ & \faTimes$^*$ & \faThumbsOUp & \faTimes & \faTimes & \faThumbsOUp & ST+VLAN & \parbox[t]{8cm}{$^*$Shortest paths are offered only within spanning trees. One spanning tree\\ per VLAN is used. Focus on \textbf{resiliece}.} \\ 

Viking~\cite{sharma2004viking} & L2 & \faThumbsUp$^*$ & \faTimes$^*$ & \faThumbsOUp & \faTimes & \faTimes$^{**}$ & \faThumbsOUp & ST+VLAN & \parbox[t]{8cm}{$^*$Shortest paths are offered only within spanning trees. One spanning tree\\ per VLAN is used. $^{**}$Viking uses elaborate load balancing, but it is static.} \\ 

TeXCP~\cite{kandula2005walking} & L3 & \faThumbsOUp & \faThumbsUp & \faThumbsOUp & \faThumbsUp & \faThumbsOUp$^*$ & \faThumbsOUp & --- & \parbox[t]{8cm}{Routing in ISP, path are computed offline,\\$^*$load balancing selects paths based on congestion and failures.} \\ 

CTMP~\cite{sohn2006congestion} & L3$^*$ & \faThumbsOUp & \faTimes & \faThumbsOUp & \faThumbsOUp & \faThumbsOUp & \faThumbsOUp & ECMP & \parbox[t]{8cm}{The scheme focuses on generating paths and on adaptive load balancing.\\It extends ECMP. $^*$Path generation is agnostic to the layer.} \\

SEATTLE~\cite{kim2008floodless} & L2 & \faThumbsOUp & \faTimes & \faTimes & \faTimes & \faTimes & \faThumbsOUp & LIS (DHT) & Packets traverse the shortest paths.\\

\makecell[l]{SPB~\cite{allan2010shortest}, TRILL~\cite{touch2009transparent}} & L2 & \faThumbsOUp & \faTimes$^*$ & \faThumbsOUp & \faTimes & \faTimes & \faThumbsOUp & --- & --- \\ 

Ethernet on Air~\cite{sampath2010ethernet} & L2 & \faThumbsOUp & \faTimes & \faThumbsUp$^{*}$ & \faTimes & \faTimes & \faThumbsOUp & LIS (DHT) & \makecell[l]{$^{*}$Multipathing is used only for \textbf{resilience}.} \\

VIRO~\cite{jain2011viro} & L2--L3 & \faThumbsOUp & \faThumbsOUp & \faThumbsUp$^{*}$ & \faTimes & \faTimes & \faThumbsOUp & LIS (DHT) & \parbox[t]{8cm}{$^*$Multipathing could be enabled by using multiple virtual networks\\ over the same underlying physical topology.} \\

\makecell[l]{MLAG~\cite{subramanian2014multi}, MC-LAG~\cite{subramanian2014multi}} & L2 & \faThumbsUp$^*$ & \faTimes$^*$ & \faThumbsUp$^{**}$ & \faTimes & \faTimes & \faThumbsOUp & --- & \makecell[l]{$^*$Not all shortest paths are enabled; $^{**}$multipathing only for \textbf{resilience}.} \\ 

\midrule
\multicolumn{10}{l}{\textbf{Related to Ethernet and TCP/IP (data centers, supercomputers):}}\\
\midrule
DCell~\cite{guo2008dcell} & L2--L3 & \faTimes & \faThumbsOUp & \faTimes & \faTimes & \faTimes & \faTimes\ (RL)$^*$ & --- & $^*$DCell comes with a \emph{specific} topology that consists of layers of routers. \\

Monsoon~\cite{greenberg2008towards} & L2, L3 & \faThumbsUp$^*$ & \faTimes$^*$ & \faThumbsUp$^{**}$ & \faTimes & \faTimes & \faTimes\ (CL)$^\dagger$ & \parbox[t]{1cm}{VLB, SR,\\ECMP} & \parbox[t]{8cm}{$^*$VLB is used in groups of edge routers. $^{**}$ECMP is used only between\\ border and access routers.} \\

Work by Al-Fares et al.~\cite{alfares2008scalable} & L3 & \faThumbsOUp & \faTimes & \faThumbsOUp & \faThumbsOUp & \faThumbsOUp & \faTimes\  (FT) & --- & --- \\

PortLand~\cite{niranjan2009portland} & L2 & \faThumbsOUp & \faTimes & \faThumbsOUp$^*$ & \faTimes & \faTimes & \faTimes\ (FT) & ECMP & --- \\ 

MOOSE~\cite{scott2009addressing} & L2 & \faThumbsOUp & \faTimes & \faThumbsUp$^*$ & \faTimes & \faTimes & \faThumbsOUp & \parbox[t]{1.8cm}{OSPF-OMP$^{**}$, LIS} & \parbox[t]{8cm}{$^*$Only a brief discussion on augmenting the frame format for multipathing.\\$^{**}$only mentioned as a possible mechanism for multipathing in MOOSE.} \\

BCube~\cite{guo2009bcube} & L2--L3 & \faThumbsOUp & \faTimes & \faThumbsOUp & \faThumbsOUp & \faTimes & \faTimes\ (RL)$^*$ & --- & $^*$BCube comes with a \emph{specific} topology that consists of layers of routers. \\

VL2~\cite{greenberg2009vl2} & L3$^*$ & \faThumbsOUp & \faTimes & \faThumbsOUp & \faTimes & \faThumbsUp$^{**}$ & \faTimes\ (CL) & \makecell[l]{LIS, VLB, ECMP} & \makecell[l]{$^*$L3 is used but L2 semantics are offered. $^{**}$TCP congestion control.} \\ 

SPAIN~\cite{mudigonda2010spain} & L2 & \faThumbsUp$^*$ & \faThumbsUp$^*$ & \faThumbsOUp & \faThumbsOUp & \faTimes & \faThumbsOUp & ST+VLAN & \parbox[t]{8cm}{$^*$SPAIN uses one ST per VLAN. Path diversity is limited by \#VLANs\\ supported in L2 switches.} \\ 

Work by Linden et al.~\cite{van2011revisiting} & L3 & \faThumbsOUp & \faTimes & \faThumbsOUp & \faThumbsUp$^*$ & \faThumbsUp$^*$ & \faThumbsOUp & ECMP & \makecell[l]{$^*$These aspects are only mentioned. The whole design extends ECMP.} \\

Work by Suchara et al.~\cite{suchara2011network} & L3 & \faThumbsOUp & \faThumbsUp$^*$ & \faThumbsOUp & \faThumbsOUp & \faThumbsUp$^{**}$ & \faThumbsOUp & --- & \parbox[t]{8cm}{$^*$Support is implicit. $^{**}$Paths are precomputed based on predicted traffic.\\ The design focuses on fault tolerance but also considers performance.} \\

PAST~\cite{stephens2012past} & L2 & \faThumbsUp$^*$ & \faThumbsUp$^*$ & \faThumbsUp$^{**}$ & \faThumbsOUp & \faTimes & \faThumbsOUp & \parbox[t]{1cm}{ST+VLAN,\\VLB} & \parbox[t]{8cm}{$^*$PAST enables \emph{either} shortest \emph{or} non-minimal paths.\\$^{**}$Limited or no multipathing.} \\ 


\makecell[l]{Shadow MACs~\cite{agarwal2014shadow}} & L2 & \faThumbsOUp & \faTimes$^*$ & \faThumbsOUp & \faTimes & \faTimes & \faThumbsOUp & --- & \makecell[l]{$^*$Non-minimal paths are mentioned only in the context of \textbf{resilience}.} \\ 

WCMP for DC~\cite{zhou2014wcmp} & L3 & \faThumbsOUp$^*$ & \faTimes & \faThumbsOUp$^*$ & \faTimes & \faTimes & \faTimes\ (MS)$^{**}$ & ECMP & \parbox[t]{8cm}{WCMP uses OpenFlow~\cite{mckeown2008openflow}. $^*$WCMP extends ECMP with hashing\\ of flows based on link capacity. $^{**}$Applicable to simple 2-stage networks.} \\

\parbox[t]{2.5cm}{Flexible fabric~\cite{jyothi2015towards}} & L3 & \faThumbsOUp & \faThumbsUp$^*$ & \faTimes$^{**}$ & \faTimes & \faTimes & \faThumbsUp$^{\text{\textdagger}}$ & SR & \parbox[t]{8cm}{$^*$Non-minimal paths considered for \textbf{resilience} only. $^{**}$Only mentioned.\\$^{\text{\textdagger}}$Main focus is placed on leaf-spine and fat trees.} \\

XPath~\cite{hu2016explicit} & L3 & \faThumbsOUp & \faThumbsUp$^{*}$ & \faThumbsOUp & \faThumbsOUp & \faThumbsUp$^{**}$ & \faThumbsOUp & --- & \makecell[l]{$^*$Unclear scaling behavior. $^{**}$XPath relies on default congestion control.} \\

\makecell[l]{Adaptive load balancing$^*$} & L3 & \faThumbsOUp & \faTimes & \faThumbsOUp & \faTimes & \faThumbsOUp & \faTimes\ (MS) & PR & $^*$Examples are DRILL~\cite{ghorbani2017drill} or DRB~\cite{cao2013per} \\ 


ECMP-VLB~\cite{kassing2017beyond} & L3 & \faThumbsOUp & \faThumbsOUp & \faThumbsOUp & \faTimes & \faThumbsOUp & \faThumbsOUp\ (XP)$^*$ & \makecell[l]{ECMP, VLB} & $^*$Focus on the Xpander network. \\




%
%
FatPaths~\cite{besta2019fatpaths} & L2--L3 & \faThumbsOUp$^*$ & \faThumbsOUp$^*$ & \faThumbsOUp & \faThumbsOUp & \faThumbsOUp & \faThumbsOUp$^{**}$ & \parbox[t]{1.2cm}{PR$^{***}$, ECMP$^\dagger$, VLAN$^\dagger$} & \parbox[t]{8cm}{$^*$Simultaneous use of shortest and non-minimal paths.\\ $^{**}$Generally applicable but main focus is on low-diameter topologies.\\ $^{***}$FatPaths sprays packets grouped in flowlets. $^\dagger$Only briefly described.} \\
\midrule
\multicolumn{10}{l}{\textbf{Related to InfiniBand and other traditionally HPC-related designs (data centers, supercomputers):}}\\
\midrule

%
\makecell[l]{Shortest path schemes$^*$} & subnet & \faTimes & \faTimes & \faThumbsUp$^{**}$ & \faTimes & \faTimes & \faThumbsUp & {D-free} & \parbox[t]{8cm}{$^*$They incl.~Min-Hop~\cite{mellanox_technologies_mellanox_2013}, (DF-)SSSP~\cite{hoefler_optimized_2009,domke-hoefler-dfsssp}, and Nue~\cite{domke_routing_2016}. \\$^{**}$Only when combined with NAA.} \\

\makecell[l]{MUD~\cite{lysne_load_2001,flich_improving_2002}} & subnet & \faThumbsUp & \faThumbsUp & \faThumbsUp & \faTimes & \faTimes & \faTimes$^*$ & {D-free} & $^*$\parbox[t]{8cm}{Original proposals disregarded IB's destination-based routing criteria;\\hence, applicability is limited without NAA.} \\

\makecell[l]{LASH-TOR~\cite{skeie_lash-tor:_2004}} & subnet & \faTimes & \faTimes & \faThumbsUp & \faTimes & \faTimes & \faTimes$^*$ & {D-free} & $^*$\parbox[t]{8cm}{Original proposals disregarded IB's destination-based routing criteria;\\hence, applicability is limited without NAA.} \\

\makecell[l]{Multi-Routing~\cite{nomura_performance_2012}} & subnet & \faThumbsUp$^*$ & \faTimes & \faThumbsOUp & \faThumbsOUp & \faThumbsUp$^{**}$ & \faThumbsUp$^*$ & --- & \parbox[t]{8cm}{$^*$Depends on \#\{network planes\} and/or selected routing schemes.\\$^{**}$Must be implemented in upper layer protocol, like MPI.} \\

\makecell[l]{Adaptive Routing~\cite{mellanox_technologies_how_2019}} & subnet & --- & --- & \faThumbsOUp & --- & \faThumbsOUp & \faThumbsUp & --- & \makecell[l]{Propriety Mellanox extension are outside of InfiniBand specification.} \\

\makecell[l]{SAR~\cite{domke_scheduling-aware_2016}} & subnet & \faTimes & \faTimes & \faTimes & \faTimes & \faThumbsUp$^*$ & \faTimes\ (FT) & NAA, {D-free} & $^*$\makecell[l]{Theoretically in Phase 2 \& 4 of 'Property Preserving Network Update'.} \\ 

\makecell[l]{PARX~\cite{domke_hyperx_2019}} & subnet & \faThumbsUp & \faThumbsUp & \faThumbsOUp & \faTimes & \faThumbsUp$^*$ & \faTimes\ (HX) & NAA, {D-free} & \makecell[l]{$^*$Implemented via upper layer protocol, e.g. modified MPI library.} \\

%
%
\midrule

\makecell[l]{Cray's Aries~\cite{aries}} & propr. & \faThumbsOUp & \faThumbsOUp & \faThumbsOUp & \faTimes & \faThumbsOUp$^*$ & \faTimes\ (DF) & UGAL, {D-free} & \parbox[t]{8cm}{$^*$Link congestion information are propagated through the network and\\ used to decide between minimal and non-minimal paths.} \\

\makecell[l]{Cray's Slingshot~\cite{slingshot}} & \makecell[l]{L3 or propr.} & \faThumbsOUp & \faThumbsOUp & \faThumbsOUp & \faTimes & \faThumbsOUp$^*$ & \faTimes\ (DF) & UGAL, {D-free} & \makecell[l]{$^*$Similar to Aries, adds endpoint congestion mitigation. } \\

Myricom's Myrinet~\cite{boden1995myrinet} & propr. & \faThumbsOUp & \faThumbsOUp & \noAnswer & \noAnswer & \noAnswer & \faThumbsOUp & SR, {D-free} & --- \\

Atos' BXI~\cite{derradji2015bxi} & propr. & \faThumbsOUp & \noAnswer & \noAnswer & \noAnswer & \faThumbsOUp & \faThumbsOUp & {D-free} & --- \\

EXTOLL's architecture~\cite{neuwirth2015scalable} & propr. & \faThumbsOUp & \noAnswer & \noAnswer & \noAnswer & \faThumbsOUp & \faThumbsOUp & & --- \\

Intel's OmniPath~\cite{birrittella2015intel} & propr. & \faThumbsOUp & \faThumbsOUp & \faThumbsUp$^*$ & \faThumbsUp$^*$ & \faThumbsOUp & \faThumbsOUp & {D-free} & \makecell[l]{$^*$No built-in support for enforcing packeting ordering across different paths} \\

Quadrics' QsNet~\cite{petrini2003performance, petrini2002quadrics} & propr. & \faThumbsOUp & \faThumbsOUp & \faThumbsUp$^*$ & \faThumbsUp$^*$ & \faThumbsOUp & \faTimes\ (FT) & SR & $^*$Unclear details on how to use multipathing in practice  \\

IBM's PERCS~\cite{ibm-percs-network} & propr. & \faThumbsOUp & \faThumbsOUp & \faThumbsOUp & \faTimes & \faThumbsUp$^*$ & \faTimes\ (DF) & UGAL, {D-free} & \makecell[l]{$^*$Routing modes can be set on a per-packet basis.} \\

\bottomrule
\end{tabular}
\vspace{-1em}
\caption{
  \ssmall
\textmd{
\textbf{
Routing architectures. Rows are sorted
chronologically and then by topology/multipathing support.}
\textbf{``Scheme used''} indicates incorporated building blocks from
Table~\ref{tab:blocks}.
\textbf{``Stack Layer''} indicates the location of a given scheme in the
TCP/IP or InfiniBand stack (cf.~\cref{sec:back_arch}).
\textbf{SP}, \textbf{NP}, \textbf{MP}, \textbf{DP}, \textbf{ALB}, and \textbf{AT}
illustrate whether a given routing scheme supports various aspects of path diversity. Specifically:
\textbf{SP}: A given scheme enables using arbitrary \textbf{shortest} paths.
\textbf{NP}: A given scheme enables using arbitrary \textbf{non-minimal} paths.
\textbf{MP}: A given scheme enables \textbf{multipathing} (between two hosts).
\textbf{DP}: A given scheme considers \textbf{disjoint} (no shared links) paths.
\textbf{ALB}: A given scheme offers \textbf{adaptive load balancing}.
\textbf{AT}: A given scheme works with an \textbf{arbitrary topology}.
%
%
%
\faThumbsOUp: A given scheme does offer a given feature. \faThumbsUp: A given scheme offers a given feature
in a limited way. \faTimes: A given scheme does not offer a given feature. $^*$Explanations in remarks.
MS, FT, CL, XP, and HX are symbols of topologies described in~\cref{sec:nettopo}. RL is a specific
type of a network called ``recursive layered'' design, described in~\cref{sec:rl}.
``\noAnswer'': Unknown.
{``\textbf{D-free}'': deadlock-free.}
}
}
%
%
\label{tab:intro}

\end{table*}

\section{Routing Protocols and Architectures}
\label{sec:architectures}

We now describe representative networking architectures, focusing on their
support for \emph{path diversity} and \emph{multipathing}\footnote{\scriptsize We encourage
participation in this survey. In case the reader possesses additional
information relevant for the contents, the authors welcome the input. We also
encourage the reader to send us any other information that they deem important,
e.g., architectures not mentioned in the current survey version.}, according to
the taxonomy described in Section~\ref{sec:taxonomy}.  Table~\ref{tab:intro}
illustrates the considered architectures and the associated
protocols.  Symbols ``\faThumbsOUp'', ``\faThumbsUp'', and ``\faTimes''
indicate that a given design offers a given feature, offers a given feature in
a limited way, and does not offer a given feature, respectively. 
%
 
We broadly group the considered designs intro three classes. First
(\cref{sec:cat_small_clusters}), we describe schemes that belong to the
Ethernet and TCP/IP landscape and were introduced for {the Internet or for
small clusters}, most often for the purpose of increasing resilience, with
performance being only secondary target. Despite the fact that these schemes
originally did not target data centers, we include them as many of these
designs were incorporated or used in some way in the data center context.
Second, we incorporate Ethernet and TCP/IP related designs that are
specifically targeted at data centers or supercomputers (\cref{sec:cat_eth}).
The last class is dedicated to designs related to InfiniBand
(\cref{sec:cat_ib}).

\subsection{Ethernet \& TCP/IP (Clusters, General Networks)}
\label{sec:cat_small_clusters}

In the first part of Table~\ref{tab:intro}, we illustrate the Ethernet and
TCP/IP schemes that are associated with small clusters and general networks.
Chronologically, the considered schemes were proposed between 1999 and 2010
(with VIRO from 2011 and MLAG from 2014 being exceptions).

Multiple Spanning Trees (MSTP)~\cite{ieee2001mstp, de2006improving} extends the
STP protocol and it enables creating and managing multiple spanning trees over
the same physical network. This is done by assigning different VLANs to
different spanning trees, and thus frames/packets belonging to different VLANs
can traverse different paths in the network.  There exist Cisco's
implementations of MSTP, for example Per-VLAN spanning tree (PVST) and
Multiple-VLAN Spanning Tree (MVST).
Table-based Hashing with Reassignments (THR)~\cite{chim2004traffic} extends
ECMP to a simple form of load balancing: it selectively reassigns some active
flows based on load sharing statistics.
Global Open Ethernet (GOE)~\cite{iwata2004global, iwate2002global} provides
virtual private network (VPN) services in metro-area networks (MANs) using
Ethernet. Its routing protocol, per-destination multiple rapid spanning tree
protocol (PD-MRSTP), combines MSTP~\cite{ieee2001mstp} (for using multiple
spanning trees for different VLANs) and RSTP~\cite{ieee2003rstp} (for quick
failure recovery).
Viking~\cite{sharma2004viking} is very similar to GOE. It also relies on MSTP
to explicitly seek faster failure recovery \emph{and} more throughput by using
a VLAN per spanning tree, which enables redundant switching paths between
endpoints. 
TeXCP~\cite{kandula2005walking} is a Traffic Engineering (TE) distributed
protocol for balancing traffic in intra-domains of ISP operations.  It focuses
on algorithms for path selection and load balancing, and briefly discusses a
suggested implementation that relies on protocols such as
RSVP-TE~\cite{awduche2001rsvp} to deploy paths in routers. TeXCP is similar to
another protocol called MATE~\cite{elwalid2001mate}.
TRansparent Interconnection of Lots of Links
(TRILL)~\cite{touch2009transparent} and Shortest Path Bridging
(SPB)~\cite{allan2010shortest} are similar schemes that both rely on link state
routing to, among others, enable multipathing based on multiple trees and ECMP.
Ethernet on Air~\cite{sampath2010ethernet} uses the approach introduced by
SEATTLE~\cite{kim2008floodless} to eliminate flooding in the switched network.
They both rely on LIS and distributed hashtables (DHTs), implemented in
switches, to map endpoints to the switches connecting these endpoints to the
network. Here, Ethernet on Air uses its DHT to construct a routing substrate in
the form of a Directed Acyclic Graph (DAG) between switches. Different paths in
this DAG can be used for multipathing. VIRO~\cite{jain2011viro} is similar in
relying on the DHT-style routing. It mentions multipathing as a possible
feature enabled by multiple virtual topologies built on top of a single
physical network.
Finally, MLAG~\cite{subramanian2014multi} and
MC-LAG~\cite{subramanian2014multi} enable multipathing through link
aggregation.

First, many of these designs enable \emph{shortest paths}, but a non-negligible
number is limited in this respect by the used spanning tree protocol (i.e., the
used shortest paths are not shortest with respect to the underlying physical
topology). A large number of protocols alleviates this with different
strategies. For example, SEATTLE, Ethernet on Air, and VIRO use DHTs that
virtualize the physical topology, enabling shortest paths. Other schemes, such
as SmartBridge~\cite{rodeheffer2000smartbridge} or
RBridges~\cite{perlman2004rbridges}, directly enhance the spanning tree
protocol (in various ways) to enable shortest paths.
Second, many protocols also support \emph{multipathing}. Two most common
mechanisms for this are either ECMP (e.g., in AMP or THR) or multiple spanning
trees combined with VLAN tagging (e.g., in MSTP or GOE). However, \emph{almost
no schemes} explicitly support \emph{non-minimal paths\footnote{\scriptsize
While schemes based on spanning trees strictly speaking enable non-minimal
paths, this is not a mechanism for path diversity per se, but limitation
dictated by the fact that the used spanning trees often do not enable shortest
paths.}, disjoint paths, or adaptive load balancing}. Yet, they \emph{all}
work on \emph{arbitrary topologies}.
All these features are mainly dictated by the purpose and origin of these
architectures and protocols. Specifically, most of them were developed with the
main goal being \emph{resilient to failures} and \emph{not} higher
performance. This explains -- for example -- almost no support for
\emph{adaptive load balancing in response to network congestion}.
Moreover, they are all restricted by the technological constraints in general
Ethernet and TCP/IP related equipment and protocols, which are historically
designed for the general Internet setting. Thus, they have to support
\emph{any} network topology. Simultaneously, many such protocols were based on
spanning trees. This dictates the nature of multipathing support in these
protocols, often using some form of multiple spanning trees (MSTP, GOE, Viking)
or ``shortcutting'' spanning trees (VIRO).

\subsection{Ethernet \& TCP/IP (Data Centers, Supercomputers)}
\label{sec:cat_eth}

The designs associated with data centers
and supercomputers are listed in the second part of Table~\ref{tab:intro}. 

\subsubsection{Multistage (Fat Tree, Clos, Leaf-Spine) Designs}

One distinctive group of architectures target multistage topologies. A common 
key feature of all these designs is multipathing based on multiple paths of
equal lengths leading via core routers (cf.~\cref{sec:nettopo}). Common
building blocks are ECMP, VLB, and PR; however, details (of how these blocks
are exactly deployed) may vary depending on, for example, the specific targeted
topology (e.g., fat tree vs.~leaf-spine), the targeted stack (e.g., bare L2 Ethernet
vs.~the L3 IP setting), or whether a given design uses off-the-shelf equipment
or rather proposes some HW modifications.
Importantly, 
these designs focus on multipathing with shortest paths because multistage
networks offer a rich supply of such paths.  They often offer some form of
load balancing.

Monsoon~\cite{greenberg2008towards} provides a hybrid L2--L3 Clos design in
which all endpoints in a datacenter form a large single L2 domain.  L2 switches
may form multiple layers, but the last two layers (access and border) consist
of L3 routers.
ECMP is used for multipathing between access and border routers. All L2 layers
use multipathing based on selecting a random intermediate switch in the
uppermost L2 layer (with VLB). To implement this, Monsoon relies on switches
that support \emph{MAC-in-MAC tunneling (encapsulation)}~\cite{mac-in-mac} so
that one may forward a frame via an intermediate switch.

PortLand~\cite{niranjan2009portland} uses fat trees and provides a complete L2
design; it simply assumes standard ECMP for multipathing.

Al-Fares et al.~\cite{alfares2008scalable} also focus on fat trees. They
provide a complete design based on L3 routing. While they only briefly mention
multipathing, they use an interesting solution for spreading traffic over core
routers.  Specifically, they propose that each router maintains a
\emph{two-level routing table}. Now, a destination address in a packet may be
matched based on its prefix (``level~1''); this matching takes place when a
packet is sent to an endpoint in the same pod. If a packet goes to a different
pod, the address hits a special entry leading to routing table ``level~2''. In
this level, matching uses the address \emph{suffix} (``right-hand'' matching).
The key observation is that, while simple prefix matching would force packets
(sent to the same subnet) to use the same core router, suffix matching enables
selecting different core routers.  The authors propose to implement such
routing tables with ternary content-addressable memories (TCAM).

VL2~\cite{greenberg2009vl2} targets Clos and provides a design in which the
infrastructure uses L3 but the services are offered L2 semantics. VL2 combines
ECMP and VLB for multipathing. To send a packet, a random core router is
selected (VLB); ECMP then is used to further spread load across available
redundant paths. Using an intermediate core router in VLB is implemented with
IP-in-IP encapsulation.

There is a large number of load balancing schemes for multistage networks.  The
majority focus on the transport layer details and are outside the scope of this
work; we outline them in Section~\ref{sec:related-aspects} and coarsely
summarize them in Table~\ref{tab:intro}. An example design,
DRB~\cite{cao2013per}, offers round-robin packet spraying and it also discusses
how to route such packets in Clos via core routers using IP-in-IP
encapsulation.

\subsubsection{General Network Designs}

There are also architectures that focus on \emph{general} topologies; some
of them are tuned for certain classes of networks but may in principle work on
any topology~\cite{besta2019fatpaths}.
In contrast to architectures for multistage networks, designs for general
networks rarely consider ECMP because it is difficult to use ECMP in a context
of a general topology, without the guarantee of a rich number of redundant
shortest paths, common in Clos or in a fat tree.
Instead, they often resort to some combination of ST and VLANs.

SPAIN~\cite{mudigonda2010spain} is an L2 architecture that focuses on using
commodity off-the-shelf switches. To enable multipathing in an arbitrary
network, SPAIN (1) precomputes a set of redundant paths for different endpoint
pairs, (2) merges these paths into trees, and (3) maps each such tree into a
separate VLAN. Different VLANs may be used for multipathing between endpoint
pairs, assuming used switches support VLANs. While SPAIN relies on TCP
congestion control for reacting to failures, it does not offer any specific
scheme for load balancing for more performance. 

MOOSE~\cite{scott2009addressing} addresses the limited scalability of Ethernet;
it simply relies on orthogonal designs such as OSPF-OMP for multipathing.

PAST~\cite{stephens2012past} is a complete L2 architecture for general
networks. Its key idea is to use a single spanning tree per endpoint.  As such,
it does not explicitly focus on ensuring multipathing between \emph{pairs} of
endpoints, instead focusing on providing path diversity at the granularity of a
destination endpoint, by enabling computing different spanning trees, depending
on bandwidth requirements, considered topology, etc.. It enables shortest
paths, but also supports VLB by offering algorithms for deriving spanning trees
where paths to the root of a tree are not necessarily minimal.  PAST relies on
ST and VLAN for implementation.

There are also works that focus on encoding a diversity of paths available in
different networks. For example, Jyothi et al.~\cite{jyothi2015towards} discuss
encoding arbitrary paths in a data center with OpenFlow to enable flexible
fabric, XPath~\cite{hu2016explicit} compresses the information of paths in a
data center so that they can be aggregated into a practical number of routing
entries, and van der Linden et al.~\cite{van2011revisiting} discuss how to
effectively enable source routing by appropriately transforming selected fields
of packet headers to ensure that the ECMP hashing will result in the desired
path selection.

Some recent architectures focus on high-performance routing in low-diameter
networks. ECMP-VLB is a simple routing scheme suggested for Xpander
topologies~\cite{kassing2017beyond} that, as the name suggests, combines the
advantages of ECMP and VLB.
Finally, FatPaths~\cite{besta2019fatpaths} targets general low-diameter
networks. It (1) divides physical links into \emph{layers} that form acyclic
directed graphs, (2) uses paths in different layers for multipathing. Packets
are sprayed over such layers using flowlets. FatPaths discusses an
implementation based on address space partitioning, VLANs, or ECMP.

\subsubsection{Recursive Networks}
\label{sec:rl}

Some architectures, besides routing, also come
with novel ``recursive'' topologies~\cite{guo2008dcell, guo2009bcube}.  The key
design choice in these architectures to obtain path diversity is to use
multiple NICs per server and 
connect servers to one another.

\subsection{InfiniBand}
\label{sec:cat_ib}

We now describe the IB landscape.  We omit a line of common routing protocols
based on shortest paths, as they are not directly related to multipathing, but
their implementations in the IB fabric manager natively support NAA; these
routings are MinHop~\cite{mellanox_technologies_mellanox_2013},
SSSP~\cite{hoefler_optimized_2009},
Deadlock-Free SSSP (DFSSSP)~\cite{domke-hoefler-dfsssp}, and a
DFSSSP variant called Nue~\cite{domke_routing_2016}.

\subsubsection{Multi-Up*/Down* (MUD) routing}

Numerous variations of Multi-Up*/Down* routing have been proposed,
e.g.,~\cite{lysne_load_2001, flich_improving_2002}, to overcome the bottlenecks
and limitations of Up*/Down*. The idea is to utilize a set of Up*/Down* spanning
trees---each starting from a different root node---and choose a path
depending on certain criteria. For example, Flich at al.~\cite{flich_improving_2002}
proposed to select two roots which either give the highest amount of non-minimal
or the highest amount of minimal paths, and then randomly select from those two
trees for each source-destination pair. Similarly, Lysne et al.~\cite{lysne_load_2001}
proposed to identify multiple root nodes (by maximizing the minimal distance
between them), and load-balance the traffic across the resulting spanning
trees to avoid the usual bottleneck near a single root. Both approaches require
NAA to work with InfiniBand.

\subsubsection{LASH-Transition Oriented Routing (LASH-TOR)}

The goal of LASH-TOR~\cite{skeie_lash-tor:_2004} is not directly path diversity,
however it is a byproduct of how the routing tries to ensure deadlock-freedom
(an essential feature in lossless networks) under resource constraints.
LASH-TOR uses the LAyered Shortest Path routing for the majority of
source-destination pairs, and Up*/Down* as fall-back when LASH would exceed
the available virtual channels. Hence, assuming NAA to separate the LASH
(minimal paths) from the Up*/Down* (potentially non-minimal path), one can
gain limited path diversity in InfiniBand.

\subsubsection{Multi-Routing}

Multi-routing can be viewed as an extension of the multi-plane designs
outlined in~\cref{sec:nettopo}. In preliminary experiments, researchers
have tried if the use of different routing algorithms on similar network planes
can have an observable performance gain~\cite{nomura_performance_2012}.
Theoretically, additionally to the increased, non-overlapping path diversity
resulting from the multi-plane design, utilizing different routing algorithms
within each plane can yield benefits for certain traffic patterns and load
balancing schemes, which would otherwise be hidden when the same routing is
used everywhere.

\subsubsection{Adaptive Routing (AR)}

For completeness, we list Mellanox's adaptive routing implementation for
InfiniBand as well, since it (theoretically) increases path diversity and
offers load balancing within the more recent Mellanox-based InfiniBand
networks~\cite{mellanox_technologies_how_2019}. However, to this date, their
technology is proprietary and outside of the IB specifications. Furthermore,
Mellanox's AR only supports a limited set up topologies (tori-like, Clos-like
and their Dragonfly variation).

\subsubsection{Scheduling-Aware Routing (SAR)}

Similar to LASH-TOR, the path diversity offered by SAR was not intended as
multipathing feature or load balancing
feature~\cite{domke_scheduling-aware_2016}.  Using NAA with $\text{LMC}=1$, SAR
employs a primary set of shortest paths, calculated with a modified DFSSSP
routing~\cite{domke-hoefler-dfsssp}, and a secondary set of paths, calculated
with the Up*/Down* routing algorithm. Whenever SAR re-routes the network to
adapt to the currently running HPC applications, the network traffic must
temporarily switch to the fixed secondary paths to avoid potential deadlocks
during the deployment of the new primary forwarding rules. Hence, during each
deployment, there is a short time frame where multipathing is intended, but
(theoretically) the message passing layer could also utilize both, the primary
and secondary paths, simultaneously, outside of the deployment window without
breaking SAR's validity.

\subsubsection{Pattern-Aware Routing for HyperX (PARX)}

PARX is the only known, and practically demonstrated, routing for InfiniBand
which intentionally enforces the generation of minimal and non-minimal paths,
and mixes the usage of both for load-balancing reasons~\cite{domke_hyperx_2019},
while still adhering to the IB specifications. The idea of this routing is an
emulation of AR capabilities with non-AR techniques/technologies to overcome
the bottlenecks on the shortest path between IB switch located in the same
dimension of the HyperX topology. PARX for a 2D HyperX, with NAA and
$\text{LMC}=2$, offers between 2 and 4 disjoint paths, and adaptively selects
minimal or non-minimal routes depending on the message size to optimize for
either message latency (with short payloads) or throughput for large messages.

\vspace{-0.5em}
\subsection{Other HPC Network Designs}

Cray's \textbf{Aries} and \textbf{Slingshot} adopt the adaptive UGAL routing to
distribute the load across the network. When using minimal paths, the packets
are sent directly to the dragonfly destination group. With non-minimal paths,
instead, packets are first minimally routed to an intermediate group, then
minimally routed to the destination group. Within a group, packets are always
minimally routed.
Routing decisions are taken on a per-packet basis. They consist in selecting a
number of minimal and non-minimal paths, evaluating the load on these paths,
and finally selecting one. The load is estimated by using link load
information propagated through the network~\cite{kim2008technology}.
Applications can select different ``biasing levels'' for the adaptive routing
(e.g., bias towards minimal routing), or disable the adaptive routing and
always use minimal or non-minimal paths.

In IBM's \textbf{PERCS}, shortest paths lengths vary between one and three hops
(i.e., route within the source supernode; reach the destination supernode;
route within the destination supernode). Non-minimal paths can be derived by
minimally-routing packets towards an intermediate supernode. The maximum
non-minimal path length is five hops. 
As pairs of supernodes can be connected by more than one link, multiple
shortest paths can exist. PERCS provides three routing modes that can be
selected by applications on a per-packet basis: non-minimal, with the
applications defining the intermediate supernode; round-robin, with the
hardware selecting among the multiple routes in a round-robin manner;
randomized (only for non-minimal paths), where the hardware randomly chooses an
intermediate supernode. 

Quadrics' \textbf{QsNet}~\cite{petrini2003performance, petrini2002quadrics} is
a source routed interconnect that enables, to some extent, multipathing between
two endpoints, and comes with adaptivity in switches. Specifically, a single
routing table (deployed in a QsNet NIC called ``Elan'') translates a processor
ID to a specification of a path in the network. Now, as QsNet enables loading
several routing tables, one could encode different paths in different routing
tables. Finally, QsNet offers hardware support for broadcasts, and for
multicasts to physically contiguous QsNet endpoints.

\ifconf
\fi

{Intel's
\textbf{OmniPath}~\mbox{\cite{birrittella2015intel}} offers two mechanisms for
multipathing between any two endpoints: different paths in the fabric or
different virtual lanes within the same physical route. However, the OmniPath
architecture itself does not prescribe specific mechanisms to select a specific
path. Moreover, it does not provide any scheme for ensuring packet ordering.
Thus, when such ordering is needed, the packets must use the same path, or the
user must provide other scheme for maintaining the right
ordering.}

Finally, the specifications of Myricom's
\textbf{Myrinet}~\cite{boden1995myrinet} or
\textbf{Open-MX}~\cite{goglin2008design}, {Atos'
\textbf{BXI}~\mbox{\cite{derradji2015bxi}}, and EXTOLL's
interconnect}~\mbox{\cite{neuwirth2015scalable}} do not disclose details on
their support for multipathing. Myrinet does use source routing and works on
arbitrary topologies.
{Both BXI and EXTOLL's design offer adaptive routing
to mitigate congestion, but it is unclear if multipathing is used.}

\ifconf
\fi


\section{Related Aspects of Networking}
\label{sec:related-aspects}

\ifconf
\textbf{Congestion control \& load balancing} are strongly related to the
transport layer (L4).  This area was extensively
surveyed~\cite{xu2016congestion}. Thus, we do not focus on these aspects and we
only mention them whenever necessary. 
Overall, such adaptive load balancing can be implemented using
flows~\cite{al2010hedera}, flowcells (fixed-sized packet
series)~\cite{he2015presto}, flowlets (variable-size packet
series)~\cite{vanini2017letflow}, and single packets~\cite{ghorbani2017drill}.  
In data centers, load balancing often focuses on flow and flowlet
based adaptivity. This is because the targeted stack is often based on TCP that
suffers performance degradation whenever packets become reordered. In contrast,
HPC networks usually use packet level adaptivity, and research focuses on
choosing good congestion signals, often with hardware
modifications~\cite{garcia2012ofar, garcia2013efficient}.

Similarly to congestion control, we exclude \textbf{flow control} from our focus,
as it is also usually implemented within L4.

Some works \textbf{analyze various properties of low-diameter topologies}, for
example path length, throughput, and bandwidth~\cite{jyothi2016measuring}. 
Such works could be used in combination with our multipathing 
analysis when developing routing protocols and architectures that
take advantage of different properties of a given topology.
\fi

\iftr
\textbf{Congestion control \& load balancing} are strongly related to
the transport layer (L4).  This area was extensively covered in surveys,
covering overall networking~\cite{matrawy2003survey, callegari2013survey,
reddy2008survey, luo2001survey, jain1996congestion, widmer2001survey}, mobile
or ad hoc environments~\cite{lochert2007survey, silva2015survey,
sergiou2014comprehensive, flora2011survey, zhao2010survey, dashkova2012survey,
ghaffari2015congestion, maheshwari2014survey}, and more recent cloud and data
center networks~\cite{xu2016congestion, shoja2014comparative,
thakur2017taxonomic, zhang2018load, xu2017survey, aruna2013survey,
ghomi2017load, shaw2014survey, noormohammadpour2017datacenter, ren2014survey,
zhang2013survey, wang2015survey, joglekar2016managing, foerster2019survey,
sreekumari2016transport, polese2019survey}.  Thus, we do not focus on these
aspects of networking and we only mention them whenever necessary.  However, as
they are related to many considered routing schemes, we cite respective works
as a reference point for the reader.
Many schemes for load balancing and congestion control were proposed in recent
years~\cite{lu2018multi, benet2018mp, park2019maxpass, olteanu2016datacenter,
perry2015fastpass, li2013openflow, mittal2015timely, cardwell2016bbr,
alizadeh2011data, he2016acdc, zhuo2016rackcc, handley2017re,
raiciu2011improving, bai2014pias, alizadeh2013pfabric, vamanan2012deadline,
lu2016sed, hwang2014deadline, montazeri2018homa, jiang2008explicit,
banavalikar2016credit, alasmar2018polyraptor, bredel2014flow,
caesar2010dynamic, huang2020tuning, katta2017clove, geng2016juggler,
zhang2017resilient}.
Such adaptive load balancing can be implemented using
flows~\cite{curtis2011mahout, rasley2014planck, sen2013localflow,
tso2013longer, benson2011microte, zhou2014wcmp, al2010hedera,
kabbani2014flowbender, hopps2000analysis}, flowcells (fixed-sized packet
series)~\cite{he2015presto}, flowlets (variable-size packet
series)~\cite{katta2016clove, alizadeh2014conga, vanini2017letflow,
katta2016hula, kandula2007dynamic}, and single packets~\cite{zats2012detail,
handley2017re, dixit2013impact, cao2013per, perry2015fastpass, zats2012detail,
raiciu2011improving, ghorbani2017drill}.  
In data centers, load balancing most often focuses on flow and flowlet
based adaptivity. This is because the targeted stack is often based on TCP that
suffers performance degradation whenever packets become reordered. In contrast,
HPC networks usually use packet level adaptivity, and research focuses on
choosing good congestion signals, often with hardware
modifications~\cite{garcia2012ofar, garcia2013efficient}.

Similarly to congestion control, we exclude \textbf{flow control} from our focus,
as it is also usually implemented within L4.

Some works analyze various \textbf{properties of low-diameter topologies}, for
example path length, throughput, and bandwidth~\cite{valadarsky2015,
kathareios2015cost, jyothi2016measuring, singla2012jellyfish,
kassing2017beyond, besta2018slim, li2018exascale, kawano2018k,
harsh2018expander, kawano2016loren, truong2016layout, flajslik2018megafly,
kawano2017layout, azizi2016hhs, truong2016distributed, al2017new}. 
Such works could be used with our multipathing 
analysis when developing routing protocols and architectures that
take advantage of different properties of a given topology.
\fi

\section{Challenges}

There are many challenges related to multipathing and path diversity support in
HPC systems and data centers.

First, we predict a rich line of future routing protocols and networking
architectures targeting recent low-diameter topologies. Some of the first
examples are the FatPaths architecture~\cite{besta2019fatpaths} or the PARX
routing~\cite{domke_hyperx_2019}. However, more research is required to
understand how to fully use the potential behind such networks, especially
considering more effective congestion control and different technological
constraints in existing networking stacks.

Moreover, little research exists into routing schemes suited specifically for
particular types of workloads, for example deep
learning~\cite{ben2019modular}, linear algebra
computations~\cite{besta2020communication, besta2017slimsell,
solomonik2017scaling, kwasniewski2019red}, graph
processing~\cite{besta2015accelerating, besta2018slim, besta2017push,
besta2018log, besta2018survey, gianinazzi2018communication, besta2019practice,
besta2019demystifying}, and other distributed workloads~\cite{besta2015active,
besta2014fault, fompi-paper} and algorithms~\cite{schweizer2015evaluating,
schmid2016high}.
{For example, as some workloads (e.g., in deep learning~\mbox{\cite{ben2019demystifying}}) have more
predicable communication patterns, one could try to gain speedups with
multipath routing based on the structural network properties that are static
or change slowly.  Contrarily, when routing data-driven workloads such as graph
computing, one could bias more aggressively towards adaptive multipathing, for
example with flowlets~\mbox{\cite{besta2019fatpaths, vanini2017letflow}}.}

\ifconf
\fi

Finally, we expect the growing importance of various schemes enabling
programmable routing and transport~\cite{arashloo2020enabling,
caulfield2018beyond}. Here, one line of research will probably heavily depend
on OpenFlow~\cite{mckeown2008openflow} and, especially,
P4~\cite{bosshart2014p4}. It is also interesting to investigate how to use
FPGAs~\cite{besta2019substream, besta2019graph, de2018transformations} or
``smart NICs''~\cite{di2019network, hoefler2017spin, caulfield2018beyond} in
the context of multipathing.

\section{Conclusion}

Developing high-performance routing protocols and networking architectures in
HPC systems and data centers is an important research area. Multipathing and
overall support for path diversity is an important part of such designs, and
specifically one of the enablers for high performance. The importance of routing is
increased by the prevalence of communication intensive workloads that put
pressure on the interconnect, such as graph analytics or deep learning. 

Many networking architectures and routing protocols have been developed. They
offer different forms of support for multipathing, they are related to
different parts of various networking stacks, and they are based on
miscellaneous classes of simple routing building blocks or design principles.
To propel research into future developments in the area of high-performance
routing, we present the first analysis and taxonomy of the rich landscape of
multipathing and path diversity support in the routing designs in
supercomputers and data centers. We identify basic building blocks, we
crystallize fundamental concepts, we list and categorize existing architectures
and protocols, and we discuss key design choices, focusing on the support for
different forms of multipathing and path diversity. Our analysis can be used by
network architects, system developers, and routing protocol designers who want
to understand how to maximize the performance of their developments in the
context of bare Ethernet, full TCP/IP, or InfiniBand and other HPC stacks.

\vspace{1em}

\footnotesize 

\textbf{Acknowledgment }
The work was supported by JSPS KAKENHI Grant Number JP19H04119.

\normalsize

\ifconf
\printbibliography
\fi

\iftr
  \bibliographystyle{abbrv}
  \bibliography{references,bibl_conf}
\fi

\ifconf

\vspace{-4em}

\begin{IEEEbiographynophoto}{\scriptsize Maciej Besta}\scriptsize
is a PhD student at ETH Zurich. His research focuses on high-performance
networking and large-scale irregular graph processing.
\end{IEEEbiographynophoto}

\vspace{-4em}
\begin{IEEEbiographynophoto}{\scriptsize Jens Domke}\scriptsize 
%
is a postdoctoral researcher of the High Performance Big Data Research Team at
the RIKEN Center for Computational Science, Japan. His research focus
is on interconnects, topologies, and routing for HPC systems, as
well as performance evaluation and optimization of parallel applications.
\ifall
is a postdoctoral researcher of the High Performance Big Data Research
Team at the RIKEN Center for Computational Science (R-CCS), Japan. He received
his doctoral degree from the Technische Universit{\"a}t Dresden, Germany, in
2017 for his work on routing algorithms and interconnects. Jens started
his career in HPC in 2008 as member of the winning team of the Student Cluster
Competition at SC08. Since then, he published several peer-reviewed journal
and conference articles. Jens contributed the DFSSSP and Nue routing algorithms
to the subnet manager of InfiniBand, and built the first large-scale HyperX
prototype at the Tokyo Institute of Technology. His research focus is on
interconnects, topologies, and routing algorithms for HPC systems, as well as
performance evaluation and optimization of parallel applications.
\fi
\end{IEEEbiographynophoto}
\vspace{-4em}
\begin{IEEEbiographynophoto}{\scriptsize Marcel Schneider}\scriptsize 
is a researcher at CERN, Geneva.
His research as a MSc student at ETH Zurich focused on 
simulation of large-scale low diameter networks.
\end{IEEEbiographynophoto}
\vspace{-4em}
\begin{IEEEbiographynophoto}{\scriptsize Timo Schneider}\scriptsize 
is a researcher at ETH Zurich. He works on high-performance
networking and heterogeneous computing.
\end{IEEEbiographynophoto}
\vspace{-4em}
\begin{IEEEbiographynophoto}{\scriptsize Marek Konieczny}\scriptsize 
is a researcher of Computer Science at AGH University of Science and Technology.
His research in the Distributed Systems Research Group focuses on 
networked and data-intensive systems.
\end{IEEEbiographynophoto}
\vspace{-4em}
\begin{IEEEbiographynophoto}{\scriptsize Salvatore Di Girolamo}\scriptsize 
is a PhD student at ETH Zurich. He works on high-performance
networking, focusing on network
offloading. 
\end{IEEEbiographynophoto}
\vspace{-4em}
\begin{IEEEbiographynophoto}{\scriptsize Ankit Singla}\scriptsize 
is an Assistant Professor at ETH Zurich, where he leads the  Network Design \&
Architecture Lab.  His research aims at understanding and improving large
modern-day networks, such as the Internet and data center networks.
\end{IEEEbiographynophoto}
\vspace{-4em}
\begin{IEEEbiographynophoto}{\scriptsize Torsten Hoefler}\scriptsize 
is a Professor at ETH Zurich, where he leads the Scalable Parallel Computing
Lab. His research aims at understanding performance of parallel computing
systems ranging from parallel computer architecture through parallel programming
to parallel algorithms.
\end{IEEEbiographynophoto}

\fi

\end{document}